\documentclass[pre,aps,preprint]{revtex4-1}
\usepackage{graphicx}
\usepackage{amsfonts}
\usepackage{amssymb}
\usepackage{amsmath}
\usepackage{textcomp}

\usepackage{color}
\usepackage{psfrag}

\usepackage{epsfig}

\begin{document}

\title{Hyperuniformity, quasi-long-range correlations, and void-space constraints in
maximally random jammed particle packings. I. Polydisperse spheres}

\author{Chase E.~Zachary}
\email{czachary@princeton.edu}
\affiliation{Department of Chemistry, Princeton University, Princeton, New Jersey 08544, USA}
\author{Yang Jiao}
\email{yjiao@princeton.edu}
\affiliation{Department of Mechanical and Aerospace Engineering, Princeton University, 
Princeton, New Jersey 08544, USA}
\author{Salvatore Torquato}
\email{torquato@princeton.edu}
\affiliation{Department of Chemistry, Department of Physics, 
Princeton Center for Theoretical Science,
Program in Applied and Computational Mathematics, and
Princeton Institute for the Science and Technology of Materials,
Princeton University, Princeton, New Jersey 08544, USA}

\begin{abstract}

Hyperuniform many-particle distributions possess a local number variance that grows
more slowly than the volume of an observation window, implying that the local 
density is effectively homogeneous beyond a few characteristic length scales. 
Previous work on maximally random strictly jammed sphere packings in three dimensions
has shown that these systems are hyperuniform and 
possess unusual quasi-long-range pair correlations 
decaying as $r^{-4}$, resulting in anomalous logarithmic growth in the number variance.
However, recent work on maximally random jammed sphere packings with a size distribution
has suggested that such quasi-long-range correlations and hyperuniformity are not universal among 
jammed hard-particle systems.  In this paper we show that such systems are indeed 
hyperuniform with signature quasi-long-range correlations
by characterizing the more general local-volume-fraction fluctuations.
We argue that the regularity 
of the void space induced by the constraints of saturation and strict jamming overcomes the local 
inhomogeneity of the disk centers to induce hyperuniformity in the medium with a linear small-wavenumber
nonanalytic behavior in the spectral density, resulting in quasi-long-range spatial correlations
scaling with $r^{-(d+1)}$ in $d$ Euclidean space dimensions. 
A numerical and analytical analysis of the pore-size distribution for a binary MRJ system in addition to 
a local characterization of the $n$-particle loops governing the void space surrounding the inclusions is presented in support of our
argument.  This paper is the 
first part of a series of two papers
 considering the relationships among hyperuniformity, jamming, and regularity of the void space in hard-particle 
packings.  
\end{abstract}

\maketitle

\section{Introduction}

Maximally random jammed (MRJ) packings of hard particles are the 
most disordered structures, according to some well-defined order metrics,
that are rigorously incompressible and nonshearable \cite{ToTrDe00}.
These systems are prototypical ``glassy'' structures in the sense that
they are structurally rigid yet lack Bragg peaks in their scattering
spectra \cite{ChLu00}.  
In this sense, the idea of the MRJ state has replaced the mathematically
ill-defined notion of random close packing \cite{ToTrDe00}.  
Nearly half-a-century ago 
these systems were thought to describe the disordered structure of 
liquids \cite{Be60}, but it is now known that three-dimensional
MRJ monodisperse sphere packings possess unusual quasi-long-range (QLR)
pair correlations decaying as $r^{-4}$ \cite{DoStTo05}.  This property
is markedly different from typical liquids, in which pair correlations
decay exponentially fast \cite{ToSt03, ZaTo09, ChLu00}.  Similar QLR behavior has 
also been observed in noninteracting spin-polarized fermionic ground
states \cite{ToScZa08, ScZaTo09}, the ground state of liquid helium \cite{ReCh67}, 
and the Harrison-Zeldovich power spectrum of the density fluctuations of the
early Universe \cite{Pe93}.  
However, for each of these examples and 
for MRJ hard sphere packings, the structural origins of these correlations 
have been heretofore unknown, even for monodisperse systems.  Furthermore,
it is an open problem to generalize these QLR correlations for MRJ 
states to polydisperse packings, in which the jamming properties are
intimately related to the size-distribution of the particles \cite{torquato2002rhm}. 

Motivated by the observation that MRJ packings are structurally rigid with a well-defined 
contact network, Torquato and Stillinger conjectured \cite{ToSt03} that all strictly jammed (i.e., mechanically
rigid),
saturated \cite{FN1} 
packings of monodisperse spheres in $d$-dimensional Euclidean space $\mathbb{R}^d$ 
are hyperuniform, meaning that infinite-wavelength local density fluctuations vanish \cite{ToSt03},
a proposition for which no counterexample has been found to date \cite{FN2}. 
This conjecture suggests that saturation and strict jamming are sufficient to induce 
hyperuniformity, albeit not necessary \cite{ToSt03}.  
Hyperuniform systems play an integral role in understanding the relationship
between fluctuations in local material properties and microstructural order 
\cite{GaJaJoLe03, ToSt03, GaTo04, DoStTo05, GaJoTo08, ZaTo09}.
These systems have applications to the large-scale structure of the Universe \cite{Pe93},
the structure and collective motion of grains in vibrated
granular media \cite{WaHa96}, the structure of living cells \cite{WaYaBaBa02}, transport through composites 
and porous media \cite{ToLa91},
the study of noise and granularity of photographic images \cite{Ba64, LuTo90b}, identifying properties of
organic coatings \cite{FiKuBi92}, and the fracture of composite materials \cite{To00}.
For microstructures consisting of ``point'' particles, one considers fluctuations in the local number density
within some observation window.  Hyperuniform point patterns possess local density fluctuations that 
asymptotically 
grow more slowly 
than the volume of the window.  Recently, the concept of hyperuniformity has been
extended to include systems composed of finite-volume inclusions of arbitrary geometries \cite{ZaTo09}; 
in these cases,
the quantity of interest is the fluctuation in the so-called local volume fraction, defined as
the fraction of the volume within an observation window covered by a given phase.  Hyperuniform 
heterogeneous media possess
local-volume-fraction fluctuations that asymptotically decay faster than the volume of the observation window, 
implying that
the local volume fraction approaches a global value beyond relatively few characteristic length scales.

\begin{figure}[!tp]
\centering
$\begin{array}{c@{\hspace{0.6cm}}c}\\
\includegraphics[width=0.45\textwidth]{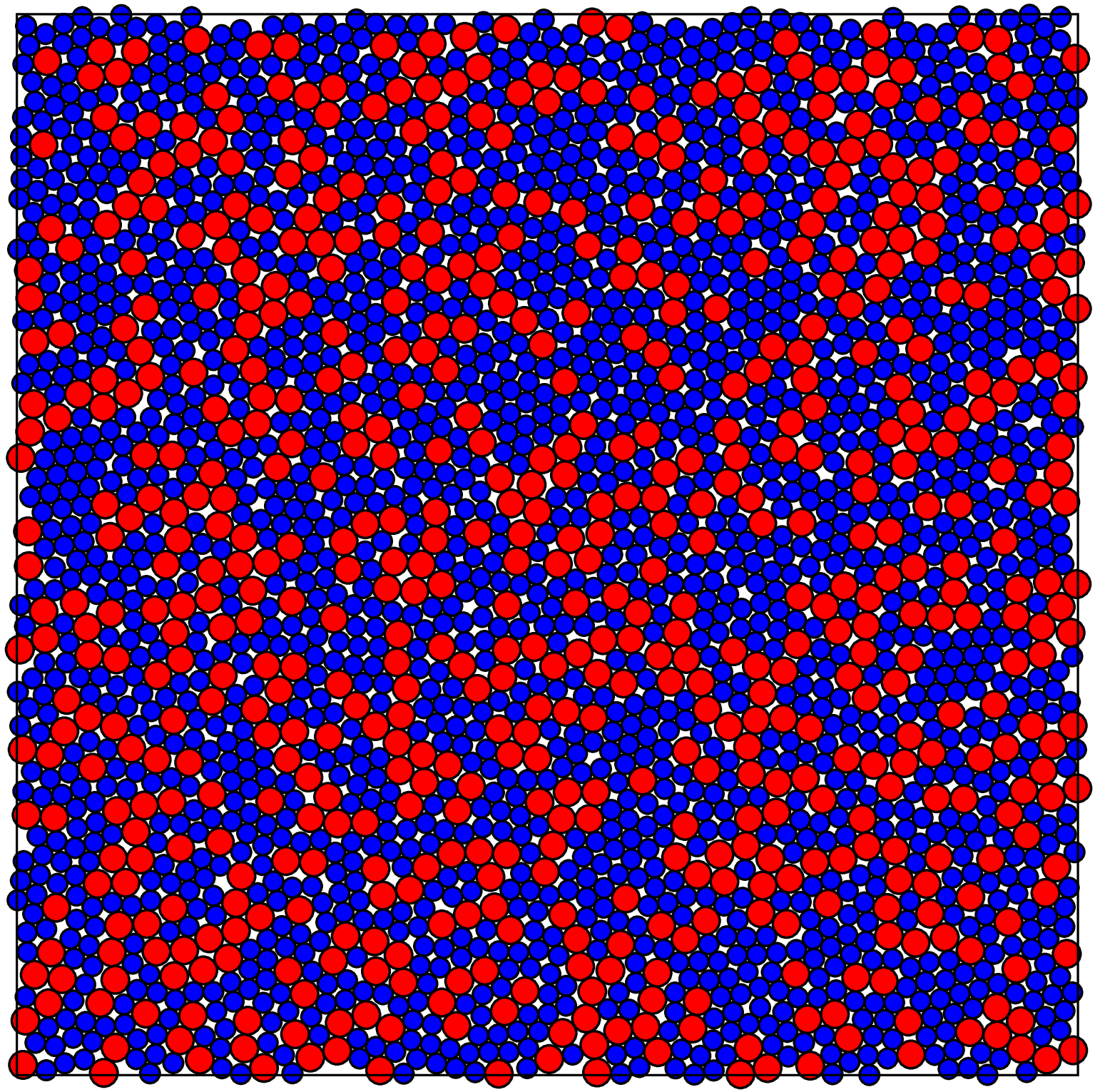}&
\includegraphics[width=0.45\textwidth]{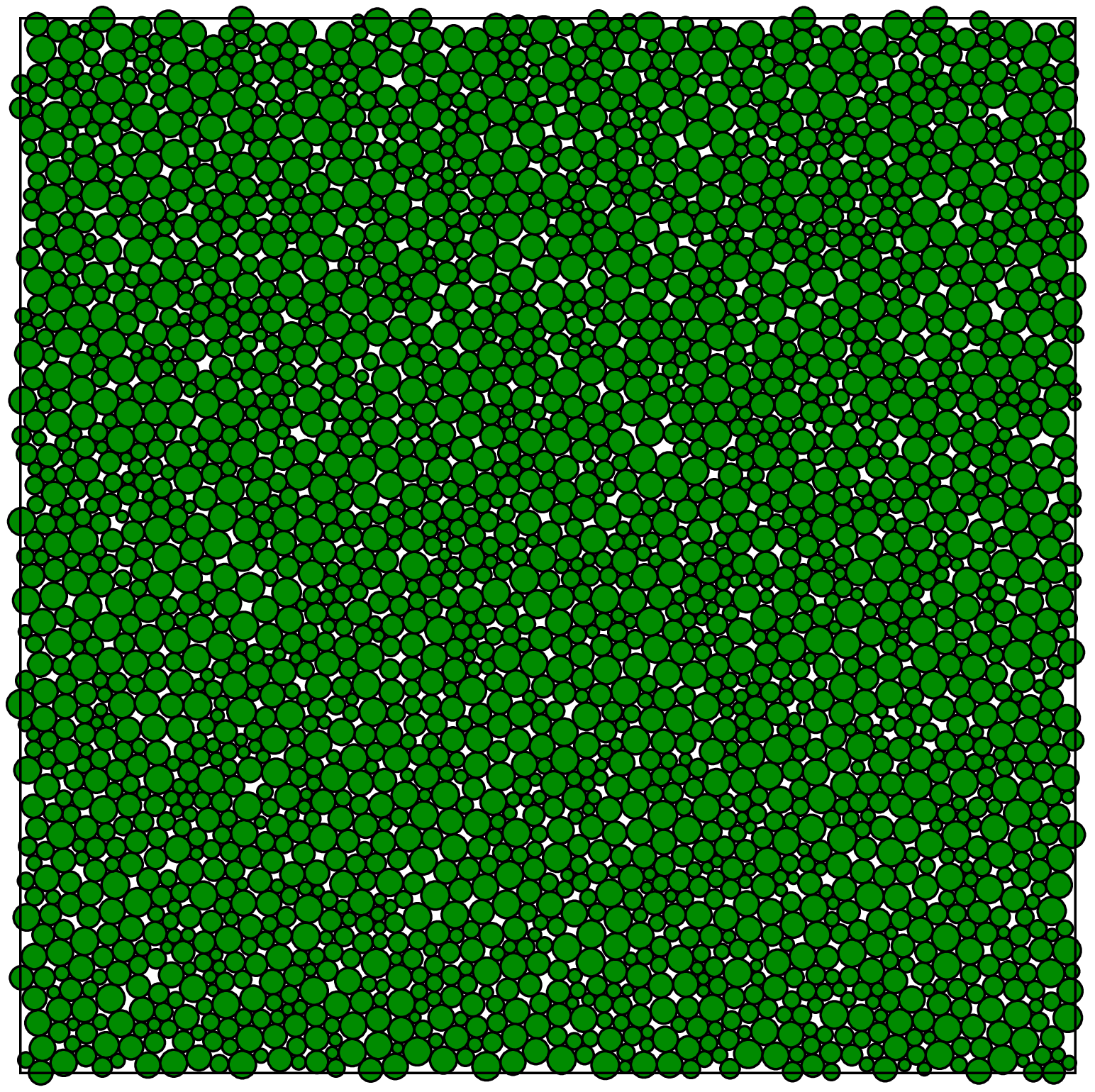}\\
\mbox{\bf (a)} & \mbox{\bf (b)}
\end{array}$
\caption{(Color online)  (a)  A binary packing of hard disks near the MRJ state. (b)
A polydisperse packing of hard disks near the MRJ state.}\label{figone}
\end{figure}

Previous work on three-dimensional (3D) MRJ monodispere sphere 
packings \cite{DoStTo05} has supported the Torquato-Stillinger conjecture
by showing that the structure factor $S(k)$, proportional to the scattering intensity, approaches zero linearly
as the wavenumber $k\rightarrow 0$, inducing a QLR power-law tail $r^{-4}$ in the 
pair correlation function $g_2(r)$.  
This behavior implies that local-number-density fluctuations grow logarithmically 
faster than the surface area of an observation window but still slower than the window volume.
In this sense, the ``degree'' of hyperuniformity in MRJ packings is minimal among all
strictly jammed saturated packings.
However, recent numerical \cite{XuCh10} and experimental \cite{KuWe10} work on polydisperse MRJ packings
has suggested that hyperuniform quasi-long-range correlations are not a universal signature of the MRJ state.
Unlike monodisperse systems, the distribution of particle sizes in a polydisperse packing 
introduces locally inhomogeneous regions
as the particles distribute themselves through space as shown in Figure \ref{figone}.
One can see in the binary packing that local clusters of small particles are distributed
near and around larger inclusions, and the result is that the
point pattern generated by the disk centers possesses local
inhomogeneities that are expected to (and indeed do) 
induce volume-order scaling within the number variance.  The situation is apparently even more 
complex for the polydisperse system in Figure \ref{figone} since the size distribution 
results in a highly inhomogeneous local structure with small particles trapped between 
larger ones with a high probability. 

These observations have raised a number of quantitatively and conceptually difficult 
questions.  First, what is the appropriate extension of the Torquato-Stillinger conjecture for 
monodisperse MRJ packings to systems with a size distribution?
Clearly, one must explicitly account for the shape information of the particles.
Second, in the event that one can 
generalize the Torquato-Stillinger conjecture, there is to date no satisfactory structural explanation 
for the \emph{linear} small-wavenumber scaling of the structure factor observed for 
3D MRJ monodisperse hard sphere packings, which indicates the presence of QLR 
pair correlations.  This extraordinarily difficult problem is tantamount to providing 
an analytical prediction of the MRJ state.  Unfortunately, no such rigorous theory 
currently exists, even for the 
considerably simpler problem of predicting 
the scalar MRJ density \cite{ToSt10}.  The presence of QLR correlations makes this problem
inherently nonlocal, and, therefore, methods that attempt to predict the MRJ state
based only on packing fraction and local criteria, such as nearest-neighbor and Voronoi
statistics, are invariably incomplete \cite{ToSt10}.  

We have presented arguments in a recent letter to suggest strongly that 
hyperuniformity and quasi-long-range pair correlations are signatures 
of saturated MRJ packings of hard particles, including binary disks, ellipses, and 
superdisks \cite{ZaJiTo11}.  
In this paper, we provide detailed evidence to show that polydisperse MRJ packings of 
hard spheres, though inhomogeneous with respect to the number variance, 
possess local-volume-fraction 
fluctuations decaying faster than the volume of
an observation window. This observation is consistent with the generalized
Torquato-Stillinger conjecture that all
strictly jammed saturated packings of spheres are hyperuniform
with respect to fluctuations in the \emph{local volume fraction}.
Our major results include:
\begin{enumerate}
\item  Infinite wavelength local number density fluctuations do not vanish for
MRJ packings of polydisperse hard spheres. 
Local-volume-fraction fluctuations provide the appropriate structural description
of these packings because they account correctly for 
the size distribution of the particles.  Importantly, our work studying local-volume-fraction
fluctuations contains previously-published results for 3D monodisperse MRJ hard sphere 
packings as a special case.
\item  Signature QLR pair correlations scaling asymptotically as $r^{-(d+1)}$ in $d$ Euclidean
dimensions are observed for all systems that we study, including binary
disks with varying size ratios and compositions and polydisperse
disk packings.  Our results suggest that these special 
correlations may be a universal feature of the MRJ state.
\item  Strict jamming places a strong constraint on the distribution of the available 
void space external to the particles such that hyperuniformity is observed when considering
local-volume-fraction fluctuations, even when the point pattern of the sphere centers is locally 
inhomogeneous.
\item  The competition between maximal randomness
and strict jamming of the packings ensures that the void-space distribution is 
sufficiently broad to induce QLR correlations between particles, thereby providing a direct 
qualitative \emph{structural} explanation for the linear small-wavenumber region of the 
generalized scattering intensity.
\end{enumerate}

\section{Background and definitions}

\subsection{Point patterns}

We consider point patterns to be realizations of stochastic point processes.
Formally, a \emph{stochastic point process} is a method of placing points in some space
(such as $\mathbb{R}^d$) according to an underlying probability distribution.
This random setting is quite general, incorporating cases in which the locations
of the points are deterministically known, such as in a Bravais lattice \cite{FN3}.

A statistically homogeneous point process is completely determined by the number density $\rho$ and the
countably infinite set (in the thermodynamic limit)
of \emph{$n$-particle correlation functions}.  The $n$-particle correlation function
$g_n (\mathbf{r}_1,\ldots, \mathbf{r}_n)$ is proportional to the probability density of finding $n$ particles in
volume elements around the positions $\mathbf{r}_1,\ldots, \mathbf{r}_n$, regardless of the positions of the
remaining particles in the system.
For an arbitrary point process, deviations of $g_n$ from unity provide a measure of the correlations among
points in the system.
Note that specifying only a finite number $M$ of the $n$-particle correlation functions defines a 
class of microstructures with degenerate $M$-particle statistics \cite{JiStTo10}.
Of particular interest is the pair correlation function $g_2$, which defines the average number of particles
surrounding a reference particle of the point process.
Closely related to the pair correlation function is the \emph{total correlation function} 
$h(\mathbf{r}) = g_2(\mathbf{r}) - 1$.
Since $g_2(r) \rightarrow 1$ as $r\rightarrow +\infty$ ($r = \lVert\mathbf{r}\rVert$) for 
isotropic, translationally invariant systems without long-range order,
it follows that $h(r)\rightarrow 0$ in this limit, meaning that $h$ is generally
integrable with a well-defined Fourier transform.

It is common in statistical mechanics when passing to reciprocal space to
consider the associated \emph{structure factor} $S(k)$, which for a translationally invariant system is defined by
\begin{equation}\label{Sdef}
S(k) = 1+\rho\hat{h}(k),
\end{equation}
where $\hat{h}$ is the Fourier transform of the total correlation function, $\rho$ is the number density, and $k = \lVert\mathbf{k}\rVert$ is the magnitude of the
reciprocal
variable to $\mathbf{r}$.
We utilize the following definition of the
Fourier transform:
\begin{equation}
\hat{f}(\mathbf{k}) = \int_{\mathbb{R}^d} f(\mathbf{r}) \exp\left(-i \mathbf{k}\cdot \mathbf{r}\right) d\mathbf{r},
\end{equation}
where $\mathbf{k}\cdot \mathbf{r} = \sum_{i=1}^d k_i r_i$ is the conventional 
Euclidean inner product of two real-valued vectors.
For radially-symmetric functions [i.e., $f(\mathbf{r}) = f(\lVert\mathbf{r}\rVert) = f(r)$], the Fourier transform may
be written 
\begin{equation}
\hat{f}(k) = (2\pi)^{d/2}\int_0^{\infty} r^{d-1} f(r) \frac{J_{(d/2)-1}(kr)}{(kr)^{(d/2)-1}} dr.
\end{equation}

\subsection{Two-phase random heterogeneous media}

Closely related to the notion of a stochastic point process is that of a two-phase random 
heterogeneous medium (or random set), 
which we define to be a domain of space $\mathcal{V} \subseteq \mathbb{R}^d$ of volume $V \leq +\infty$
that is composed of two regions:
the phase 1 region $\mathcal{V}_1$ of volume fraction $\phi_1$ 
and the phase 2 region $\mathcal{V}_2$ of volume fraction $\phi_2$ \cite{LuTo90}.
The statistical properties of each phase of the system are specified by the countably 
infinite set of \emph{$n$-point probability functions} $S_n^{(i)}$, which are defined by
\cite{ToSt82, ToSt83, torquato2002rhm, St95}
\begin{equation}\label{Sndef}
S_n^{(i)}(\mathbf{r}_1, \ldots, \mathbf{r}_n) = \left\langle\prod_{i=1}^n I^{(i)}(\mathbf{r}_i)\right\rangle,
\end{equation}
where $I^{(i)}$ is the indicator function for phase $i$
\begin{equation}
I^{(i)}(\mathbf{x}) = \begin{cases}
1, & \mathbf{x} \in \mathcal{V}_i\\
0, & \text{else}.
\end{cases}
\end{equation}
The function $S_n$ defines the probability of finding $n$ points at
positions $\mathbf{r}_1, \ldots, \mathbf{r}_n$ all within the same phase.  

Upon subtracting the long-range
behavior from $S_2$, one obtains the autocovariance function 
$\chi(\mathbf{r}) = S_2(\mathbf{r})-\phi^2$, which is generally integrable.
It is important to recognize that the autocovariance function is 
independent of the choice of reference phase, meaning that it is a \emph{global} descriptor of correlations
within the system.  This property will play a particularly important role 
in this paper when we consider the relationship between local-volume-fraction fluctuations
and the void space between inclusions in the microstructure.
The analog of the structure factor in this context is the so-called \emph{spectral density}, 
which is the Fourier transform $\hat{\chi}$ of the autocovariance function
\cite{FN4}.  
The autocovariance function obeys the bounds \cite{torquato2002rhm}
\begin{equation}
-\text{min}\{(1-\phi)^2, \phi^2\} \leq \chi(\mathbf{r}) \leq (1-\phi)\phi,
\end{equation}
where $\phi$ is the volume fraction of an arbitrary reference phase.
We remark that it is an open problem to identify additional necessary and sufficient conditions 
that the autocovariance 
function must satisfy in order to correspond to a binary stochastic 
process \cite{torquato2002rhm, To99, To06, JiStTo07, Qu08}.

\subsection{Sphere packings and a categorization of jamming}

A \emph{sphere packing} is a collection of non-overlapping
spheres in $d$-dimensional Euclidean space $\mathbb{R}^d$. The
\emph{packing density} $\phi$ (equivalent to the volume fraction
of the particle phase) is defined as the fraction of space
covered by the spheres, which may be polydisperse. An important
characteristic of a packing is its degree of randomness (or the
antithesis, order), which reflects nontrivial information of the
packing structure. The degree of randomness (order) can be
quantified by a set of \textit{order metrics} $\psi$
\cite{ToTrDe00}. It is an open and challenging problem to identify good
order metrics, but it has recently been proposed that hyperuniformity 
is itself a measure of order on large length scales \cite{ToSt03, ZaTo09}.

One method of classifying sphere packings involves characterizing
the extent to which particles are \emph{jammed}.
Torquato and Stillinger \cite{ToSt01, ToDoSt03} have provided a precise
definition of the term \emph{jamming} and have proposed a hierarchical
classification scheme for sphere packings by invoking the notions of \emph{local},
\emph{collective}, and \emph{strict} jamming. 
A packing is {\it locally} jammed if no particle in
the system can be translated while fixing the
positions of all other particles.
A {\it collectively} jammed packing is locally jammed
such that no subset
of spheres can simultaneously be continuously displaced without
moving its members out of contact both with one another
and with the remainder set. A packing is {\it strictly} jammed if it
is collectively
jammed and if all globally uniform volume nonincreasing deformations of
the system boundary are disallowed
by the impenetrability constraints. The reader is referred to Ref. \cite{ToSt01}
for further details.

As previously mentioned, the \textit{maximally random jammed} (MRJ) state is defined as the
most disordered jammed packing in a given jamming category (i.e.,
locally, collectively, or strictly jammed) \cite{ToTrDe00}. The
MRJ state is well-defined for a given jamming category and choice
of order metric, and it has recently supplanted the ill-defined
random close packed (RCP) state \cite{ToTrDe00}. In this paper, we focus on
maximally random strictly jammed polydisperse sphere packings in 
$\mathbb{R}^d$.

\subsection{Hyperuniformity in point processes:  local number density fluctuations}

A \emph{hyperuniform point process} has the property that the variance in the number of 
points in an observation window $\Omega$ grows more slowly than the volume
of that window.  In the case of a spherical observation window, this definition 
implies that the local number variance $\sigma^2_N(R)$ grows more slowly than $R^d$
in $d$ dimensions, where $R$ is the radius of the observation window.  
Torquato and Stillinger \cite{ToSt03} have provided
an exact expression for the local number variance
of a statistically homogeneous point process in a spherical observation window
\begin{equation}\label{numvar}
\sigma^2_N(R) = \rho v(R) \left[1+\rho \int_{\mathbb{R}^d} h(\mathbf{r}) \alpha(r; R) d\mathbf{r}\right],
\end{equation}
where $R$ is the radius of the observation window, $v(R)$ is the volume of the 
window, and $\alpha(r; R)$ is the
so-called \emph{scaled intersection volume}.  The latter quantity is geometrically 
defined as the volume of space occupied by the intersection
of two spheres of radius $R$ separated by a distance $r$ and normalized by the volume of a sphere $v(R)$.
Exact expressions for $\alpha(r; R)$ in arbitrary dimensions have been given by Torquato and Stillinger
\cite{ToSt06}.

It is convenient to introduce a dimensionless density $\phi$, which need not correspond
to the volume fraction, according to
\begin{equation}
\phi = \rho v(D/2) = \frac{\rho \pi^{d/2} D^d}{2^d\Gamma(1+d/2)},
\end{equation}
where $D$ is a characteristic length scale of the system 
(e.g., the mean nearest-neighbor distance between points).
The number variance
admits the following asymptotic scaling \cite{ToSt03}:
\begin{equation}\label{numasymp}
\sigma_N^2(R) = 2^d \phi\left\{A_N\left(\frac{R}{D}\right)^d 
+ B_N\left(\frac{R}{D}\right)^{d-1} + o\left[\left(\frac{R}{D}\right)^{d-1}\right]\right\},
\end{equation}
where $o(x)$ denotes all terms of order less than $x$.  This result is valid for all periodic point patterns (including
lattices), quasicrystals that possess Bragg peaks, and disordered systems in which
the pair correlation function $g_2$ decays to unity exponentially fast \cite{ToSt03}.
Explicit forms for the asymptotic coefficients $A_N$ and $B_N$ are given by \cite{ToSt03}
\begin{align}
A_N &= 1+\rho\int_{\mathbb{R}^d} h(\mathbf{r}) d\mathbf{r}= \lim_{\lVert\mathbf{k}\rVert\rightarrow 0} 
S(\mathbf{k})\label{An}\\
B_N &= -\frac{\rho \kappa(d)}{D}\int_{\mathbb{R}^d} h(\mathbf{r}) \lVert\mathbf{r}\rVert d\mathbf{r}\label{Bn},
\end{align}
where $\kappa(d) = \Gamma(1+d/2)/\{\pi^{1/2}\Gamma[(d+1)/2]\}$.  

Any system with $A_N = 0$ satisfies the requirements for hyperuniformity.  
Although the expansion \eqref{numasymp} will hold for all periodic and quasiperiodic point patterns
with Bragg peaks, this behavior is not generally true for disordered hyperuniform systems.
For example, it is known that if the total correlation function $h \sim r^{-(d+1)}$ 
for large $r$ [$S(k) \sim k$ for small $k$], then $\sigma^2_N(R) 
\sim (a_0\ln R + a_1) R^{d-1}$ \cite{ToScZa08}.
Such behavior occurs in maximally random jammed monodisperse sphere packings in 
three dimensions \cite{DoStTo05}
and noninteracting spin-polarized fermion ground states \cite{ToScZa08, ScZaTo09}.  
Other examples of ``anomalous'' local density fluctuations
have been characterized by Zachary and Torquato \cite{ZaTo09}.

\subsection{Hyperuniformity in two-phase random heterogeneous media:  local-volume-fraction 
fluctuations}

In order to define hyperuniformity for heterogeneous media we introduce the \emph{local volume fraction} $\tau_i(\mathbf{x})$ of phase $i$ according to
\begin{equation}\label{vfracone}
\tau_i(\mathbf{x}; R) = \frac{1}{v(R)}\int I^{(i)}(\mathbf{z}) w(\mathbf{z}-\mathbf{x}; R) d\mathbf{z},
\end{equation}
where $v(R)$ is the volume of the observation window and $w$ is the corresponding indicator function.
Using this definition, the variance $\sigma_{\tau}^2(R)$ in the local volume fraction is given by
\begin{equation}\label{vfraceight}
\sigma_{\tau}^2(R) = \frac{1}{v(R)} \int_{\mathbb{R}^d} \chi(\mathbf{r}) \alpha(r; R) d\mathbf{r},
\end{equation}
which is independent of the choice of reference phase.
The variance in the local volume fraction admits the asymptotic expansion \cite{ZaTo09}
\begin{align}
\sigma^2_{\tau} &= \frac{\rho}{2^d\phi}\left\{A_{\tau} \left(\frac{D}{R}\right)^d + B_{\tau} \left(\frac{D}{R}\right)^{d+1}
+ o\left[\left(\frac{D}{R}\right)^{d+1}\right]\right\}\label{tauasymp}\\
A_{\tau} &= \int_{\mathbb{R}^d} \chi(\mathbf{r}) d\mathbf{r} = \lim_{\lVert\mathbf{k}\rVert\rightarrow 0} 
\hat{\chi}(\mathbf{k})\label{atau}\\
B_{\tau} &= -\frac{\kappa(d)}{D} \int_{\mathbb{R}^d} \lVert \mathbf{r}\rVert \chi(\mathbf{r}) d\mathbf{r}\label{btau}.
\end{align}
The coefficients $A_{\tau}$ and $B_{\tau}$ in \eqref{atau} and \eqref{btau} control the asymptotic 
scaling of the fluctuations in the local volume fraction.
It then follows that $\sigma_{\tau}^2$ decays faster than $R^{-d}$ as $R\rightarrow +\infty$ for 
those systems such that
\begin{equation}\label{vfracnineteen}
\lim_{\lVert\mathbf{k}\rVert\rightarrow 0}\hat{\chi}(\mathbf{k}) = 0,
\end{equation}
which defines a hyperuniform two-phase random heterogeneous medium \cite{ZaTo09}.  

\subsection{The effect of polydispersity on local fluctuations}

Here we consider how the presence of polydispersity in a sphere packing affects the 
fluctuations in the local number density 
and local volume fraction.
For a packing of polydisperse spheres, the distribution of sphere radii is determined 
by a probability density $f(R)$, and the fundamental statistical
descriptors of the medium therefore involve averages $\langle\cdot\rangle_R$ \cite{FN5}
over the distribution of sphere sizes \cite{torquato2002rhm}.  
For a packing of polydisperse spheres with $M$ distinct radii, the density function $f(R)$ takes the form
\begin{equation}
f(R) = \sum_{i=1}^M \gamma_i \delta(R-R_i),
\end{equation}
where $\gamma_i = N_i/N$ is the mole fraction of species $i$.  As an example, we will 
continually refer in this paper to the average particle diameter $\langle D\rangle_R$
of a polydisperse hard-disk packing:
\begin{equation}
\langle D\rangle_R = \sum_{i=1}^M  \gamma_i D_i,
\end{equation}
where $D_i$ is the diameter of species $i$.

The autocovariance function $\chi(\mathbf{r})$ for a heterogeneous medium consisting of 
impenetrable polydisperse spheres is given 
by \cite{GiBlSt90, LuTo91, torquato2002rhm}
\begin{equation}\label{chipolyD}
\chi(\mathbf{r}) = \rho\langle v_{\text{int}}(\mathbf{r}; R)\rangle_R + \rho^2 \int \left\langle 
h(\mathbf{x}; R_1, R_2) v_{\text{int}}(\mathbf{r}-\mathbf{x}; R_1, R_2)
\right\rangle_{R_1, R_2} d\mathbf{x},
\end{equation}
which implies that
\begin{align}
A_{\tau} &= \int \chi(\mathbf{r}) d\mathbf{r}\\
 &= \rho\langle v^2(R)\rangle_R +\rho^2 \left\langle v(R_1) v(R_2)\int h(\mathbf{r}; R_1, R_2) 
 d\mathbf{r}\right\rangle_{R_1, R_2}\label{Atdep}.
\end{align}
Unlike for monodisperse sphere packings \cite{ZaTo09}, the result \eqref{Atdep} 
shows that it is generally not possible to separate the shape information of the inclusions
from the details of the point pattern generated by the sphere centers.  This 
observation suggests that for polydisperse microstructures, hyperuniformity
of the underlying point pattern does not induce hyperuniformity with 
respect to local-volume-fraction fluctuations; conversely, it is also possible to
find heterogeneous media for which $A_{\tau} = 0$ but $A_N > 0$.  

We note that the first term contributing to $\chi(\mathbf{r})$ in \eqref{chipolyD} can be interpreted 
as the probability of finding two points, separated by a displacement $\mathbf{r}$, in a single 
particle of the packing.  The second term is therefore related to the probability of finding the points
in two different particles.  It is clear that only this latter term, containing the pair correlation function, 
can contribute to the linear small-wavenumber region of the spectral density, and it is therefore
responsible, albeit in a highly nontrivial way, for the onset of QLR correlations in MRJ packings.  

\section{Local density and volume fraction fluctuations in polydisperse MRJ packings}

\subsection{Generation of MRJ polydisperse hard disk packings}

Motivated by the generalized conjecture that all strictly jammed packings of 
$d$-dimensional spheres are hyperuniform with respect to local-volume-fraction 
fluctuations, we have generated several packings of binary disks (2D)
near the MRJ state using a modified Lubachevsky-Stillinger (LS) packing algorithm 
\cite{LuSt90, LuStPi91, DoToSt05, DoToSt05b, DoToSt05c}, wherein particles 
with a fixed size ratio undergo event-driven molecular dynamics 
while simultaneously increasing in size according to some prescribed growth rate.  
The initial growth rate used in our simulations is 
$\gamma = 0.01,$ but near the jamming point, a much smaller growth rate 
$\gamma = 10^{-6}$ is used to establish well-defined interparticle contacts.
The Lubachevsky-Stillinger algorithm has been shown to produce MRJ
packings with these parameters consistent with the positively-correlated 
translational and orientational order metrics \cite{JiStTo11, TrToDe00}.
Our statistics for binary packings are averaged over 50 configurations of 
10000 particles; for polydisperse packings we have averaged over 10 configurations
of 10000 particles.  Our results have been compared to systems of up to $10^6$ particles
to verify invariance of the statistics with respect to system size.  

For this study, we have 
chosen a particle size ratio $\beta = R_{\text{large}}/R_{\text{small}} = 1.4$ and mole 
fractions $\gamma_{\text{small}} = 0.75$ and $\gamma_{\text{large}} = 0.25$; 
however, the focus of this study
is to elucidate a universal property of polydisperse MRJ packings, and our results are 
expected to apply for a range of size ratios, size distributions, and 
mole fractions (see Figure \ref{figtwo}) \cite{FN6}.
For the remainder of this paper, we therefore focus on the binary case with the disclaimer 
that our results are expected to apply for general polydisperse MRJ packings.  
This algorithm has been used in previous studies of 3D monodisperse 
sphere packings near the MRJ state \cite{DoToSt05}.
The resulting packings in this study have a final volume fraction $\phi \approx 0.8475$, 
which is below the the close-packed density $\phi_{\text{cp}} = \sqrt{3}\pi/6$.  
Figure \ref{figone} (Section I) 
provides a typical realization of a binary packing; note that the particle distribution is saturated and nonperiodic.

\subsection{Hyperuniformity and local-volume-fraction fluctuations}

We have calculated both the structure 
factor $S(k)$ and the spectral density $\hat{\chi}(k)$ for the 
binary MRJ disk packings using discrete Fourier transforms of the local density $\rho(\mathbf{r})$ and 
indicator function $I(\mathbf{r})$ of the particle phase of the packing.  Specifically,
\begin{align}
S(\mathbf{k}) &= \frac{\left\lvert \sum_{j=1}^N\exp\left(-i \mathbf{k}\cdot \mathbf{r}_j\right)\right\rvert^2}{N}
\qquad (\mathbf{k}\neq\mathbf{0})\\
\hat{\chi}(\mathbf{k}) &= \frac{\left\lvert \sum_{j=1}^N \exp\left(-i \mathbf{k}\cdot \mathbf{r}_j\right) 
\hat{m}(\mathbf{k}; R_j)
\right\rvert^2}{V} \qquad (\mathbf{k}\neq \mathbf{0})\label{chik},
\end{align}
where
\begin{equation}
\hat{m}(\mathbf{k}; R) \equiv \int_{\mathbb{R}^d} 
\exp(-i \mathbf{k}\cdot\mathbf{r}) \Theta(R-\lVert\mathbf{r}\rVert) d\mathbf{r}
\end{equation}
is the Fourier transform of the indicator function for a $d$-dimensional sphere of radius $R$ \cite{FN7}.  
Note that the 
shape of the enclosure (defined by a set of basis vectors $\{\mathbf{e}_i\}$) restricts the wavevectors
such that $\mathbf{k}\cdot\mathbf{e}_i = 2\pi n$ for all $i$, where $n \in \mathbb{Z}$.  Since the zero wavevector
is removed from the spectrum in the expressions above, one must define
\begin{equation}
\hat{\chi}(\mathbf{0}) \equiv \lim_{N, V\rightarrow+\infty} \hat{\chi}(\lVert\mathbf{k}\rVert = k_{\text{min}}),
\end{equation}
with a similar expression for $S(\mathbf{0})$.
The limit here is taken at constant number density $\rho = N/V$, and $k_{\text{min}}$ is the smallest
computable wavevector as determined by the shape of the boundary.   To obtain radially-symmetric forms
of the structure factor and spectral density, we radially average over all wavevectors with equal magnitude.

\begin{figure}[!tp]
\centering
$\begin{array}{c@{\hspace{0.5cm}}c}\\
\includegraphics[width=0.3\textwidth]{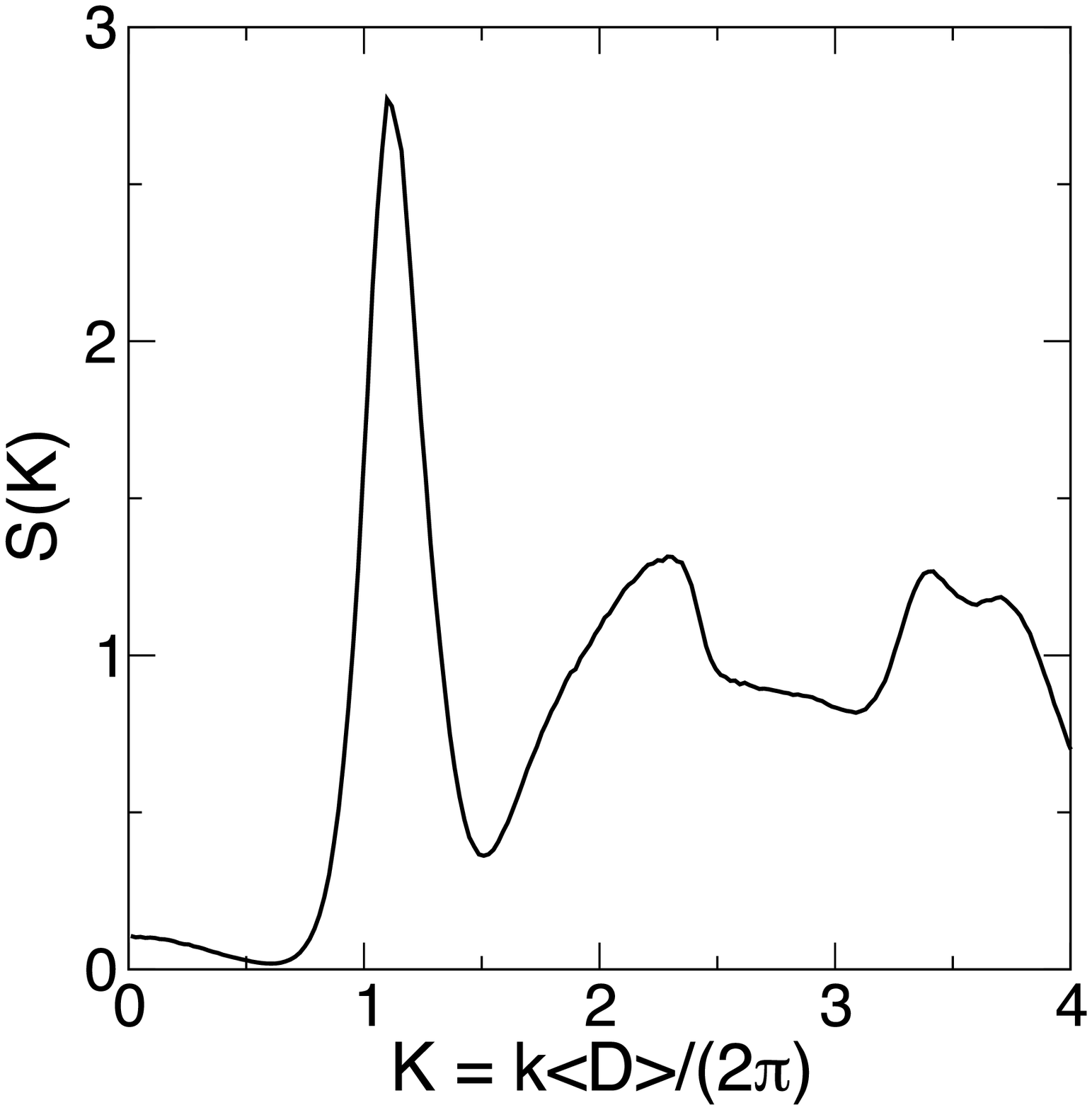} &
\includegraphics[width=0.35\textwidth]{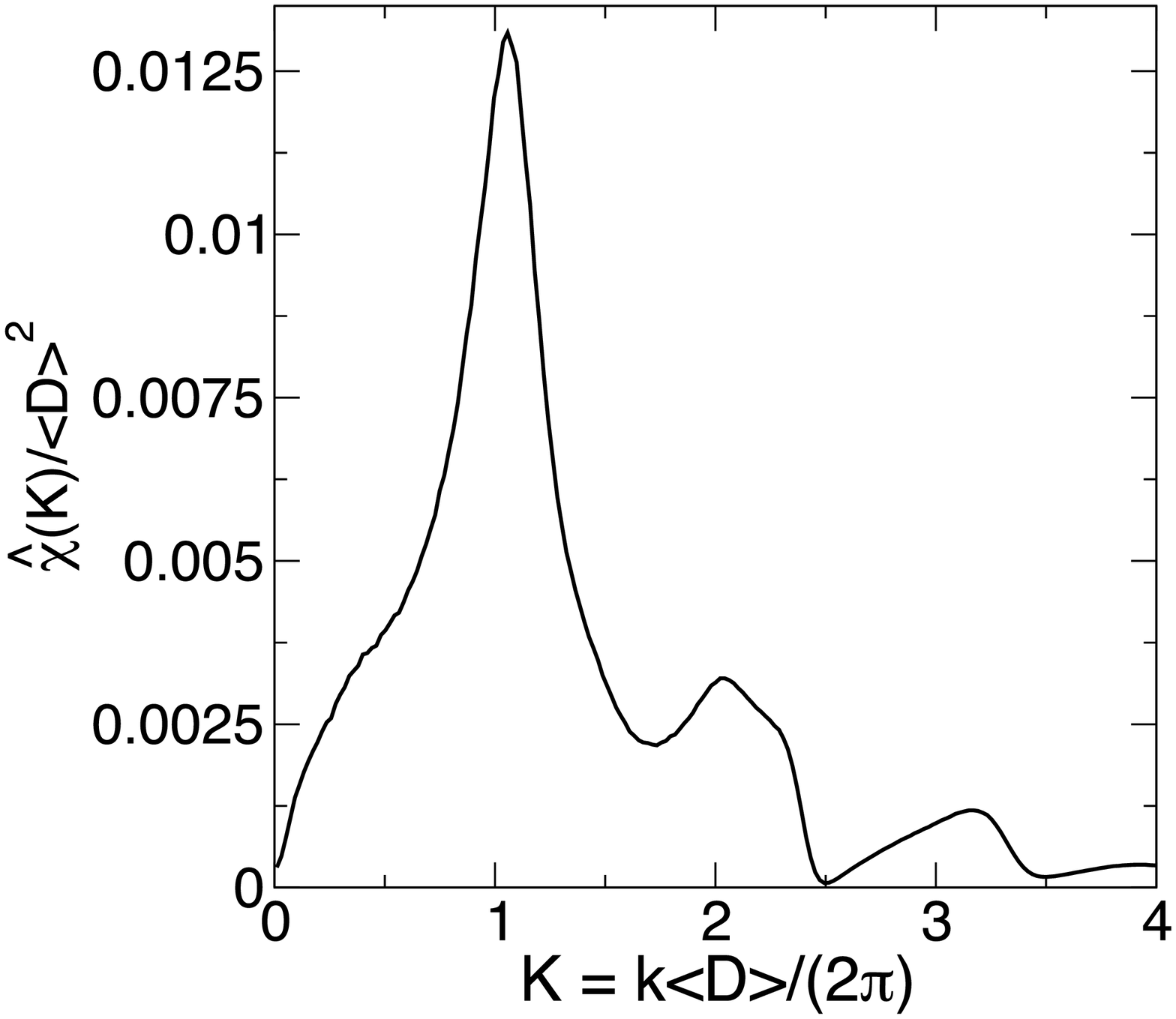} \\
\mbox{\bf (a)} & \mbox{\bf (b)}\\
\includegraphics[width=0.35\textwidth]{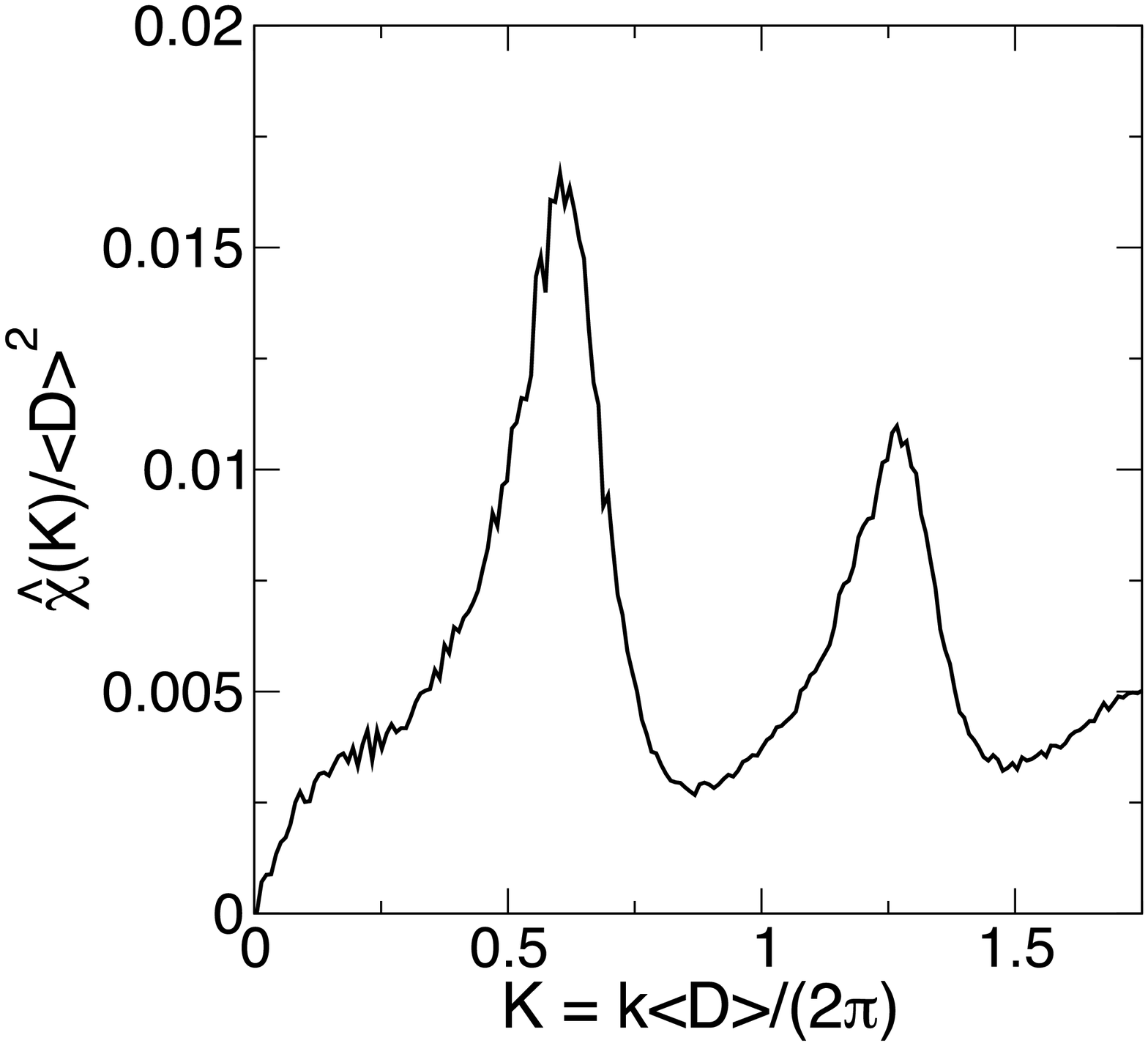} &
\includegraphics[width=0.35\textwidth]{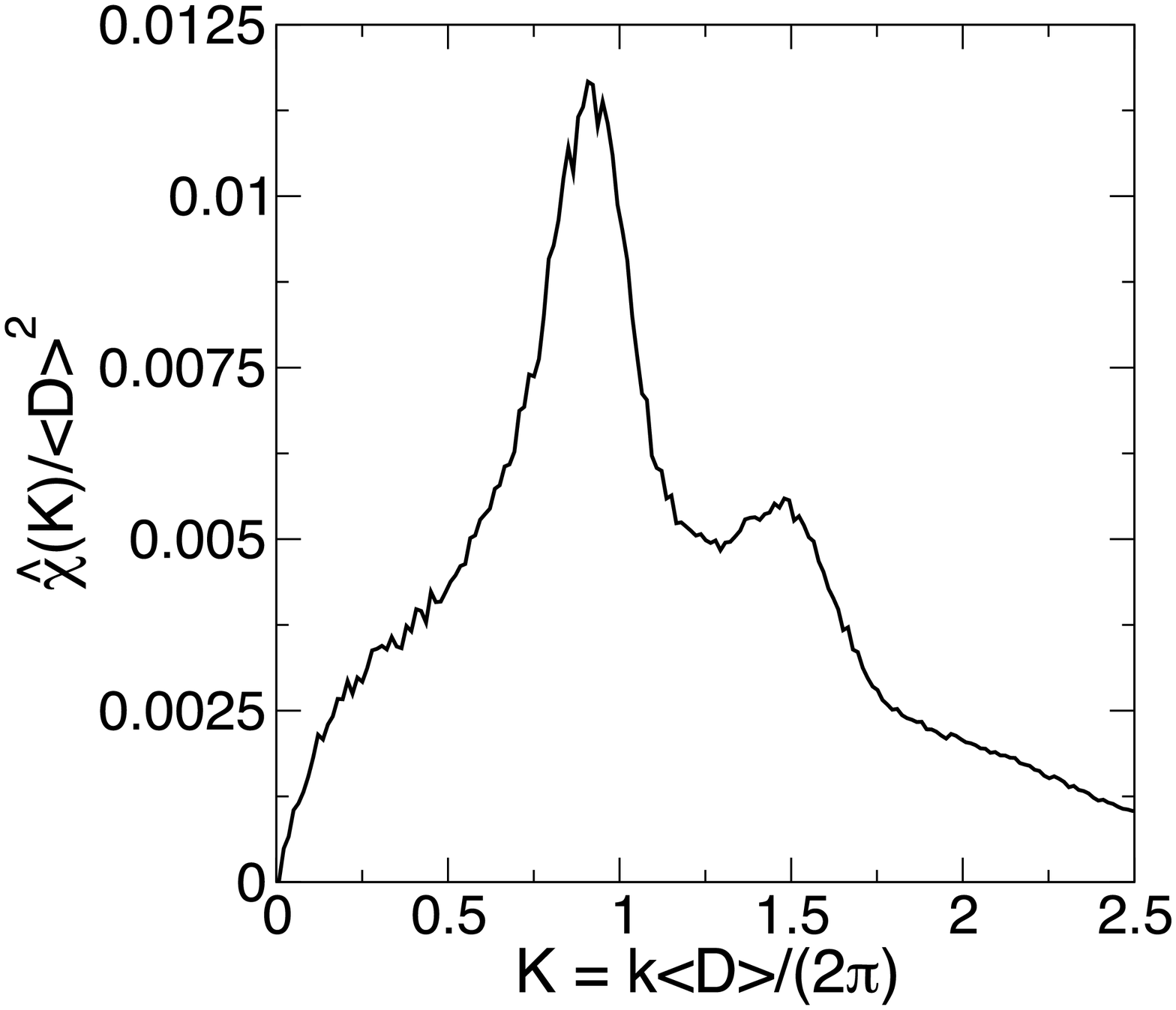} \\
\mbox{\bf (c)} & \mbox{\bf (d)}
\end{array}$
\caption{(a) Structure factor $S(k)$ for a binary packing of disks near the MRJ state.  
(b)  The corresponding spectral density $\hat{\chi}(k)$.  Note that 
infinite-wavelength local density fluctuations are not suppressed unlike local volume fraction 
fluctuations on the equivalent length scale.  
(c)  Spectral density
for particle concentrations $\gamma_{\text{small}} = 2/3$ and $\gamma_{\text{large}} = 1/3$ 
with size ratio $\beta = 2.5$.  
Hyperuniformity of the packing is unaffected
by changes in these parameters as expected by the Torquato-Stillinger conjecture.  (d)  
Spectral density 
for a polydisperse MRJ packing of disks with a uniform distribution of radii in the 
interval $[R_{\text{min}}, R_{\text{max}}]$.}\label{figtwo}
\end{figure}

\begin{figure}[!tp]
\centering
$\begin{array}{c@{\hspace{0.6cm}}c}\\
\includegraphics[width=0.45\textwidth]{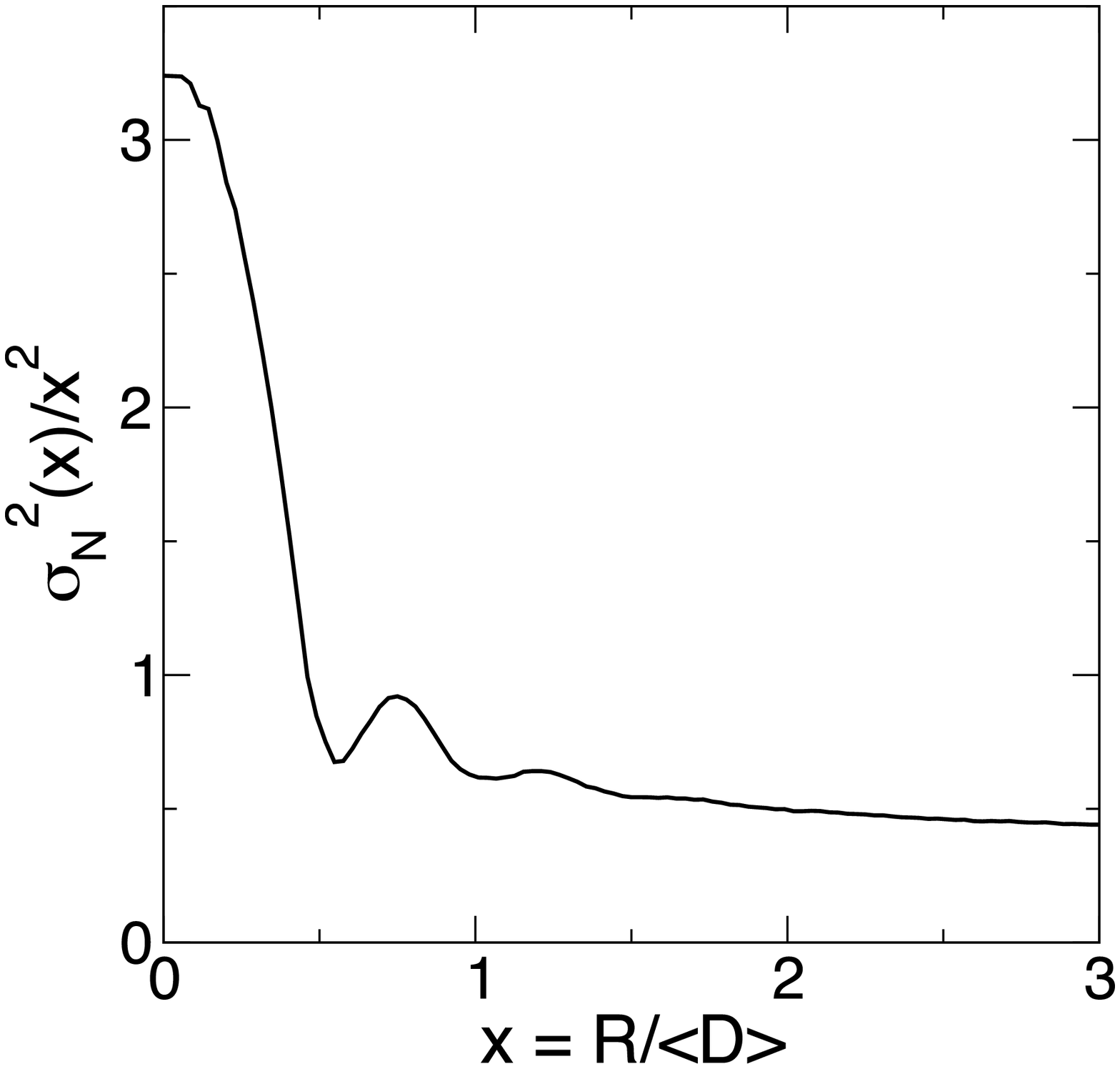} &
\includegraphics[width=0.45\textwidth]{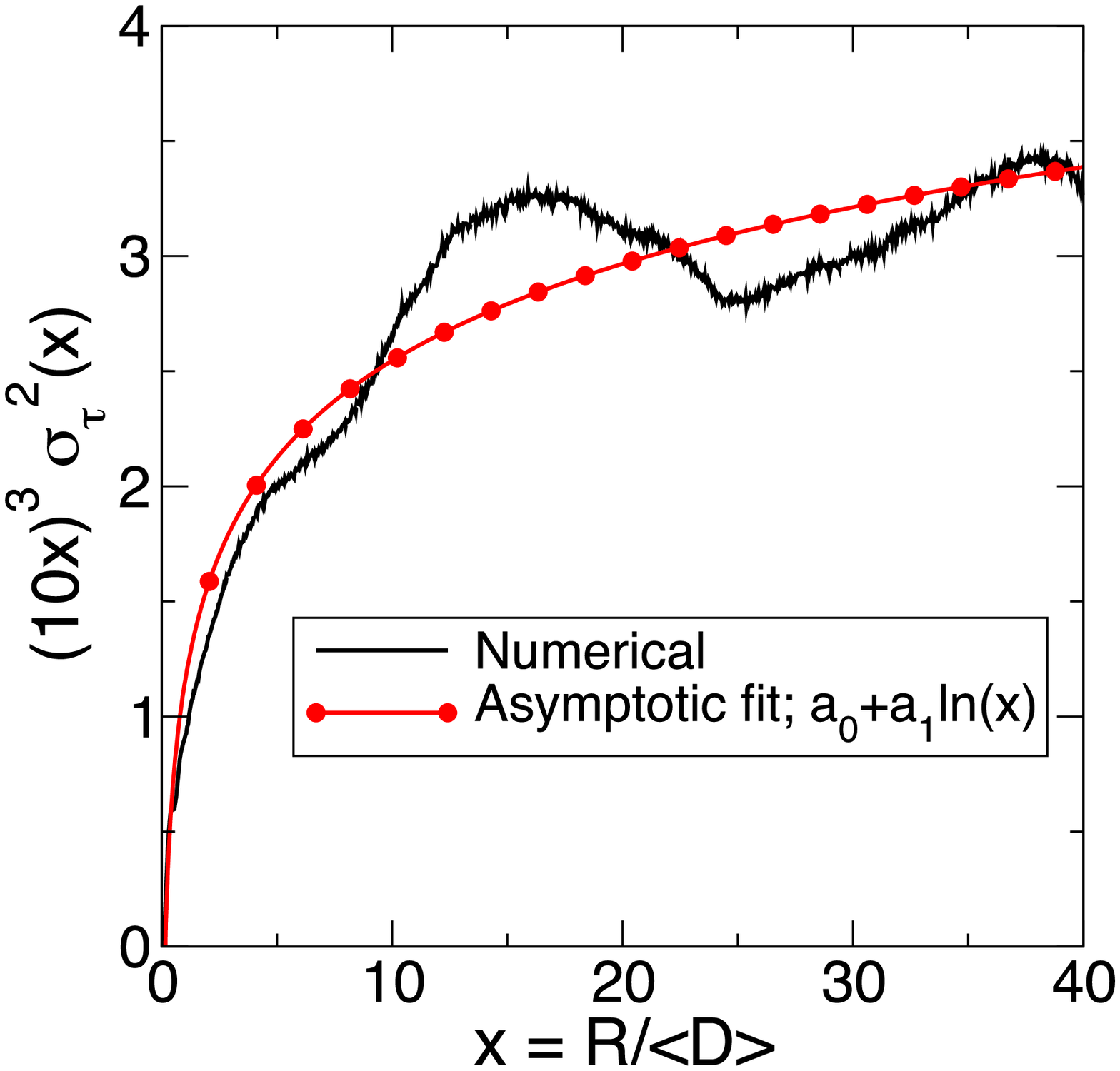} \\
\mbox{\bf (a)} & \mbox{\bf (b)} 
\end{array}$
\caption{(Color online)  (a)  Asymptotic fluctuations in the local number density for the MRJ binary 
disk packing.  (b) 
Asymptotic fluctuations in the local volume fraction for the binary MRJ packing, 
demonstrating that the system is indeed
hyperuniform.}\label{MRJnum}
\end{figure}

Our results for the structure factors and spectral densities of the 
binary and polydisperse MRJ hard disk packings are shown in Figure
\ref{figtwo}.  
The structure factors for these systems lacks Bragg peaks, reflecting the absence of long-range order.
Of particular importance is the behavior of the structure 
factor near the origin.  A fit of the small-$k$ region ($k \lesssim 0.5$) with a third-order 
polynomial suggests $S(0) \approx 0.104 > 0$, meaning that the point pattern 
generated by the disk centers does not possess 
vanishing infinite-wavelength local number density
fluctuations, an observation verified by direct calculation of the number 
variance in Figure \ref{MRJnum}.  
Note that the local number variance asymptotically
scales with the volume of the observation window.
This behavior differs from MRJ packings of 
\emph{monodisperse} spheres in three dimensions \cite{DoToSt05}, 
where the structure factor
decays linearly to zero for small wavenumbers.  
Bidisperse distributions of 
disks are inherently \emph{inhomogeneous} with respect to the 
locations of the particle centers.
For the systems studied here, the large concentration of small particles results in local 
clusters that are essentially close-packed. However, the 
introduction of large particles into the system generates
effective grain boundaries between these local clusters, breaking 
the uniformity of the underlying point pattern.    

\begin{figure}
\centering
\includegraphics[width=0.5\textwidth]{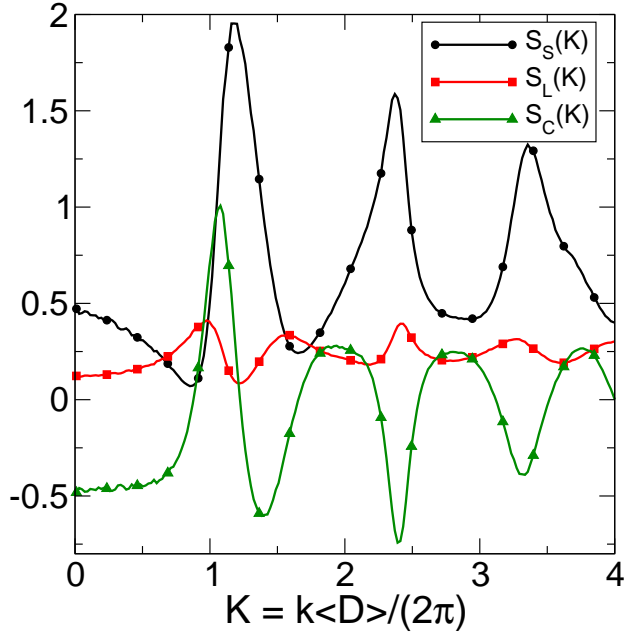}
\caption{(Color online)  
Partial contributions $S_S(k)$, $S_L(k)$, and $S_C(k)$ to the structure factor of the binary MRJ disk packings.}\label{pSk}
\end{figure}

We note that the structure factor for our binary packing can be decomposed as
\begin{equation}
S(k) = S_{S}(k) + S_{L}(k) + S_{C}(k),
\end{equation}
where
\begin{align}
S_{S}(k) &= \frac{\left\lvert \sum_{j=1}^{N_S} \exp(-i \mathbf{k}\cdot \mathbf{r}_j)\right\rvert^2}{N}\\
S_{L}(k) &= \frac{\left\lvert \sum_{j=1}^{N_L} \exp(-i \mathbf{k}\cdot \mathbf{x}_j)\right\rvert^2}{N}\\
S_{C}(k) &= 2\text{Re}\left\{\frac{\left[\sum_{j=1}^{N_S} \exp(-i \mathbf{k}\cdot \mathbf{r}_j)\right]\left[\sum_{\ell=1}^{N_L} \exp(i\mathbf{k}\cdot\mathbf{x}_\ell)\right]}{N}\right\}
\end{align}
are the particle structure factors incorporating small-small, large-large, and small-large correlations, respectively.  Note that $N_S$ is the number of small
particles with positions $\{\mathbf{r}_i\}$, and $N_L$ is the number of large particles with positions $\{\mathbf{x}_i\}$.  These partial contributions to the 
structure factor are shown in Figure \ref{pSk}.
None of the partial contributions to the structure factor possess a vanishing small-wavenumber region, implying that the local density fluctuations 
of the small and large particles each scale with the volume of an observation window as does the covariance.  
These observations clearly demonstrate that information contained in the structure factor
is not sufficient to characterize the packings because it neglects the details of the particle shapes.

A direct calculation of the spectral density for the binary MRJ system shows markedly 
different behavior from the structure factor as seen in Figure \ref{figtwo}.  
One notices that
infinite-wavelength local-volume-fraction fluctuations are 
suppressed by the system.
We have fit the small-$k$ region of the 
spectral density (in units of $\langle D\rangle_R^2$) using a third-order 
polynomial of the form $a_0 + a_1 K + a_2 K^2 + a_3 K^3$ and have found 
$a_0 = (1.0 \pm 0.2) \times 10^{-5},$ strongly suggesting 
that this system is hyperuniform with respect to local-volume-fraction fluctuations.  
We have verified this claim by directly calculating the variance in the local volume fraction, 
shown in Figure \ref{MRJnum}.  
The local-volume-fraction fluctuations 
decay faster than the volume of an observation window and logarithmically slower than the surface area, 
consistent with the presence of QLR correlations.

\subsection{Properties of the spectral density of binary MRJ hard disk packings}

\begin{figure}[!tp]
\centering
\includegraphics[width=0.50\textwidth]{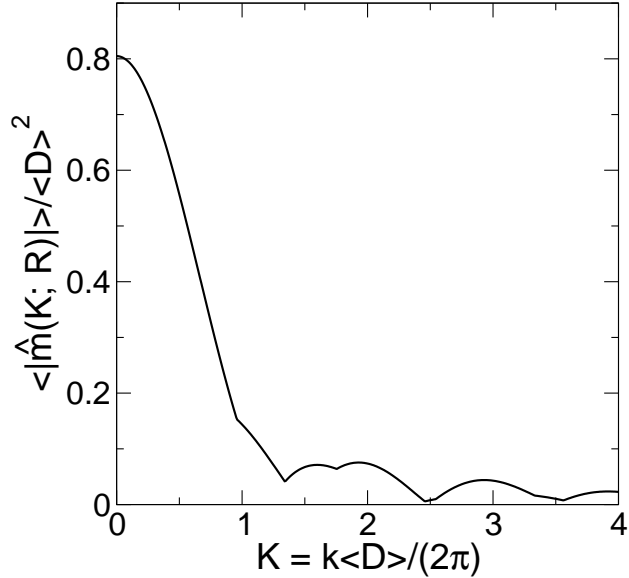}
\caption{The average magnitude $\langle \lvert\hat{m}(k; R)\rvert\rangle_R$ of the spatial contribution to the spectral density.}\label{mag}
\end{figure}

The spectral density is dominated by a peak with wavelength just below $\langle D\rangle_R$, approximately corresponding to the small-particle diameter.  
This observation is consistent with the
microstructure of the medium, which contains several regions of almost close-packed clusters of small particles.  In order to further understand the behavior of 
the spectral density, it is important to 
recognize that both the local distribution of inclusions and the shape information of the particles contribute to $\hat{\chi}$.  However, the relative influence of each component on 
the spectral density varies throughout the spectrum.  To characterize this effect, we recall the 
representation \eqref{chik} of the spectral density, which admits the following 
upper bound:
\begin{align}
\hat{\chi}(\mathbf{k}) &\leq N\rho \left(\langle \lvert \hat{m}(k; R)\rvert\rangle_R\right)^2\label{chiUL}.
\end{align}
Figure \ref{mag} plots the average $\langle \lvert \hat{m}(k; R)\rvert\rangle_R$, which according to \eqref{chiUL} controls the upper bound
on the spectral density. 
Note that for small wavenumbers [$k\langle D\rangle_R/(2\pi) \lesssim 2$] the bound \eqref{chiUL} places only weak constraints on the spectral density, implying that 
this region of the spectrum is controlled by information in the local structure of the heterogeneous medium.  
We emphasize that the shape information of the particles must still be included at small wavenumbers
to account for vanishing infinite-wavelength local-volume-fraction fluctuations, but, as we show
below, such geometric information only provides the appropriate weights for 
the partial contributions to the structure factor 
to induce hyperuniformity.
In constrast, the large-wavenumber region of the 
spectrum closely follows the behavior of the upper bound \eqref{chiUL}, meaning that the length scale imposed by the decoration of the particle centers,
here chosen to be disks, controls the spectral density.  We emphasize that this portion of the spectrum arises from the shape information 
of the inclusions themselves and is almost entirely independent of the distribution of particles in the system.  Specifically, the particle indicator function $m(r; R)$, 
which has compact support $[0, R]$, possesses a Fourier representation $\hat{m}(k; R)$ that is both long-ranged and has an intrinsic period, which for a binary packing
leads to interference effects in the large-wavenumber region.  

\begin{figure}[!tp]
\centering
\includegraphics[width=0.50\textwidth]{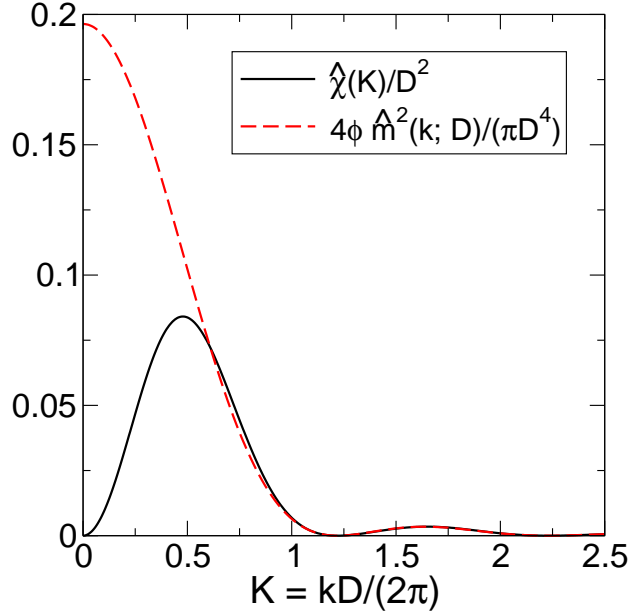}
\caption{(Color online)  Spectral 
density and corresponding shape contribution for a 
microstructure generated by the step-function $g_2$ process.  Note that the shape 
contribution controls the spectral density for large wavenumbers.}\label{stepchi}
\end{figure}

Similar behavior arises in monodisperse systems, where the spectral density is exactly given 
$\hat{\chi}(k) = \rho \hat{m}^2(k; R) S(k)$, 
and one can rigorously separate information contained in the point pattern generated by the sphere centers from the shape information of the inclusions. Figure \ref{stepchi}
shows the spectral density and shape contribution for a system of impenetrable disks with pair correlation function $g_2(r) = \Theta(D-r)$.  Note that the bound \eqref{chiUL}
only applies when $S(k)$ is at a maximum; for clarity we have omitted the associated scaling factor on the shape contribution.
As with the binary MRJ packing, this ``step-function process'' possesses a spectral density that for small wavenumbers depends signficantly on the 
distribution of sphere centers; the large-wavenumber region is almost exactly equal to the shape contribution since the structure factor 
approaches an asymptotic value of unity.  Although for binary systems polydispersity precludes the direct separation of the shape information 
from the underlying point pattern, these examples show that qualitatively these components continue to affect only specific portions of the spectral density.

\begin{figure}[!tp]
\centering
\includegraphics[width=0.50\textwidth]{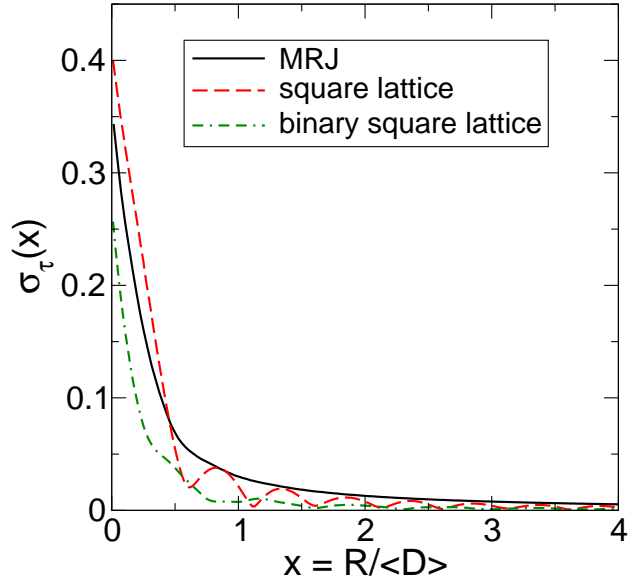}
\caption{(Color online)  Standard deviation of the local volume fraction for the close-packed 
binary square and square lattices with the corresponding 
result for the MRJ binary packing.}\label{MRJvar}
\end{figure}

The dominance of the shape information for large wavenumbers has the surprising 
effect of almost completely suppressing fluctuations for $K = k\langle D\rangle_R/(2\pi) 
\approx 2.5$,
thereby suggesting that local volume fraction fluctuations essentially vanish 
on this length scale.  In actuality, the nonuniformity of the microstructure
precludes a complete extinction of the variance in the local volume fraction; 
nevertheless, the local-volume-fraction variance (Figure \ref{MRJvar}) undergoes a sharp change in slope 
near this length scale, highlighting a transition to the 
``geometrically-controlled'' region of the spectral density.  Physically, this behavior implies
that observation windows with radii given by the appropriate wavelength 
capture the effective pore size surrounding the inclusions in such a way that the 
medium is essentially homogeneous on this \emph{local} scale.  
This observation suggests that the void space surrounding the particles plays a central role 
in determining the spatial statistics of a heterogeneous system.     

\subsection{Probing the origin of QLR pair correlations}

\begin{figure}[!tp]
\centering
$\begin{array}{c@{\hspace{0.6cm}}c}\\
\includegraphics[width=0.45\textwidth]{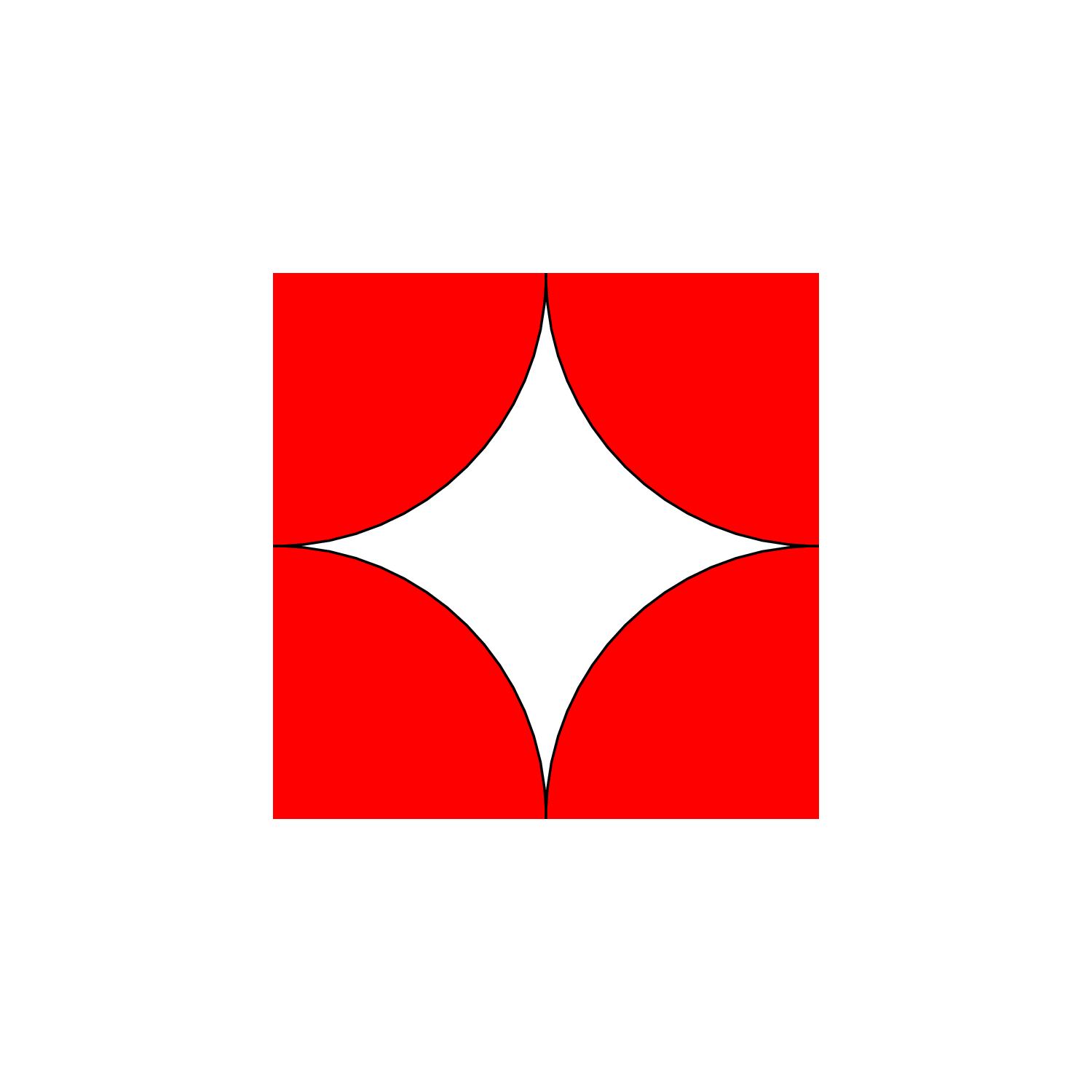}&
\includegraphics[width=0.45\textwidth]{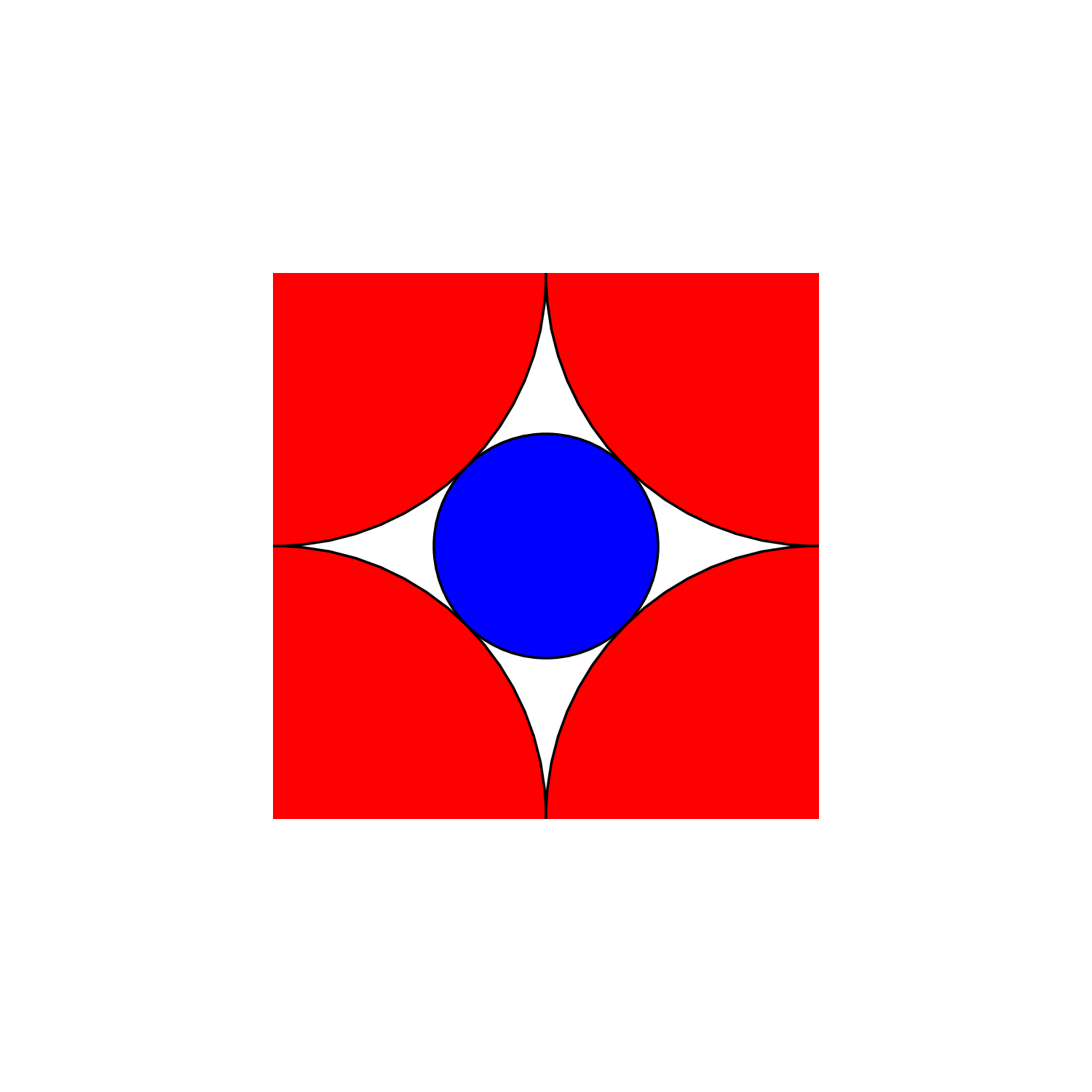}\\
\mbox{\bf (a)} & \mbox{\bf (b)}
\end{array}$
\caption{(Color online)  
(a)  Unit cell for the square lattice at close-packing. (b)  Unit cell for a 
close-packed binary variant of the square lattice.}\label{ZbinZ}
\end{figure}

It is important to note that the small-$k$ behavior of the 
spectral density for the binary MRJ packing results in an ``anomalous'' asymptotic scaling of 
the variance in the local volume fraction.  Specifically, 
our results suggest that the spectral density is nonanalytic at the origin with 
an expansion $\hat{\chi}(k) \sim a_1 k + \mathcal{O}(k^2)$ as $k \rightarrow 0$, 
implying that the autocovariance function
exhibits quasi-long-range behavior and scales with $r^{-3}$ [$r^{-(d+1)}$ in $d$ dimensions].  
Equivalent behavior has been observed for the 
structure factor 
of MRJ monodisperse sphere packings in three dimensions \cite{DoStTo05}, and the number 
variance in that case has been shown to 
scale according to:
\begin{equation}
\sigma^2_N(R) \sim (b_0 + b_1\ln R) R^2 + \mathcal{O}(R).
\end{equation}
A related scaling must also hold for the variance in the local volume fraction of our binary MRJ packings; 
specifically:
\begin{equation}
\sigma^2_{\tau}(R) \sim \frac{(c_0 + c_1\ln R)}{R^3} + \mathcal{O}(R^{-4}).
\end{equation}
One can directly see the effect of this behavior in Figure \ref{MRJvar}, where the variance in the local 
volume fraction for the MRJ packings is compared to 
the square lattice packing and its binary variant, the unit cells for which are shown in Figure \ref{ZbinZ}.  
Note that local volume fraction 
fluctuations in the MRJ packings decay more slowly than in either of the periodic systems; indeed, 
as the size distribution of the void space 
becomes smaller and more uniform,
local-volume-fraction fluctuations are more rapidly suppressed on the global scale 
of the microstructure.  Since the integer packing is not even strictly jammed,
this comparison suggests that strict jamming is neither a necessary nor a strong determinant of 
hyperuniformity in heterogeneous media.

Identifying the origin of the linear small-wavenumber region of the spectral density is an open 
problem that must be 
related to the structural features of 
the MRJ state.  This problem is particularly difficult for our polydisperse packings since the 
underlying point pattern 
generated by the particle centroids possesses
non-vanishing infinite-wavelength 
local number density fluctuations.  
However, the spectral density $\hat{\chi}(k)$ can be expressed in terms of the partial structure factors as
\begin{equation}
\hat{\chi}(k) = \rho \hat{m}^2(k; R_S) S_S(k) + \rho \hat{m}^2(k; R_L) S_L(k) + \rho \hat{m}(k; R_S) 
\hat{m}(k; R_L) S_C(k),
\end{equation}
where $R_S$ and $R_L$ are the radii of the small and large particles, respectively.  
This result follows directly from an expansion of \eqref{chik}.
As $k \rightarrow 0$, we therefore find
\begin{equation}\label{qform}
\hat{\chi}(0) = \rho v^2(R_S) S_S(0) + \rho v^2(R_L) S_L(0) + \rho v(R_S) v(R_L) S_C(0),
\end{equation}
implying that the particle volumes provide the appropriate \emph{weights} to properly balance the small- and large-particle variances 
with the covariance between the particles.  Additionally, since $\hat{m}(k; R)$ possesses no linear term in its small-wavenumber Taylor
expansion, it follows that the appearance a linear small-wavenumber region in the spectral density, and therefore QLR correlations, \emph{must} 
involve an appropriate superposition of linear contributions in the partial structure factors.  

This observation suggests that quasi-long-range 
correlations in MRJ packings arise from the competing effects of strict 
jamming and maximal randomness.  Incompressibility of the structure
from strict jamming implies that particles should be correlated over 
several characteristic length scales; indeed, along the equilibrium branch
of the binary hard-disk phase diagram, one expects that correlations 
become fully long-ranged at the close-packed density, corresponding to 
phase-separated lattice structures \cite{DoStTo07}.  However, maximal 
randomness interferes with this long-range order, resulting in the
apparent $r^{-(d+1)}$ asymptotic scaling in the pair correlation function.  
For our binary packings, this scaling is encoded in the partial 
pair correlations of the structure, which, after appropriate weighting with 
the shape information of the particles, induces hyperuniformity. 

We emphasize that a complete explanation for the appearance of the 
linear small-wavenumber region of the spectral density is
intractable because the problem is inherently non-local due to hyperuniformity
and the presence of QLR correlations.  Furthermore, it has recently been 
established that decreasing the 
exponential scaling of the small-wavenumber is associated with greater 
disorder within a many-particle system, potentially
even inducing clustering among the particles \cite{ZaTo10}, an effect clearly 
inconsistent with strict jamming.  More generally, 
hyperuniformity is associated with an effective interparticle repulsion that 
attempts to evenly distribute the particles throughout
space, and the length scale of this repulsion increases with increasing exponential 
scaling of the small-wavenumber region
of the spectral density \cite{ZaTo10}.  The conditions of saturation and impenetrability 
in an MRJ packing induce structural 
order that apparently competes with the constraint of maximal randomness to
minimize the scaling of the small-wavenumber region of the spectral density to its 
smallest integer value, an effect that we argue
is physically tied to the void-space distribution of the system.  

\begin{figure}[!t]
\centering
\includegraphics[width=0.65\textwidth]{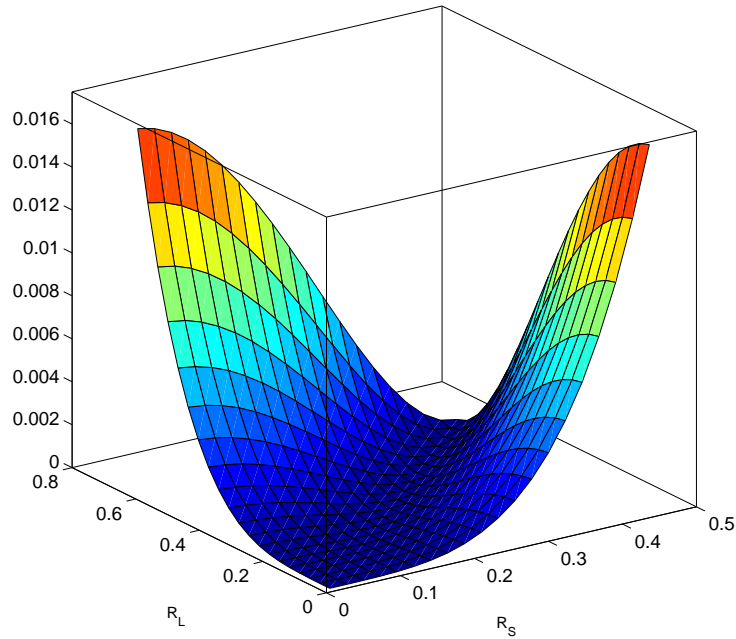}
\caption{(Color online)  The quadratic form \eqref{qform}, equal to the small-wavenumber limit of the spectral density $\hat{\chi}(0)$, as a function of the radii of the particles in the packing.  A uniform 
decrease in the particle radii defines a basin of hyperuniform systems.}\label{qformfig}
\end{figure}
\begin{figure}[!t]
\centering
\includegraphics[width=0.50\textwidth]{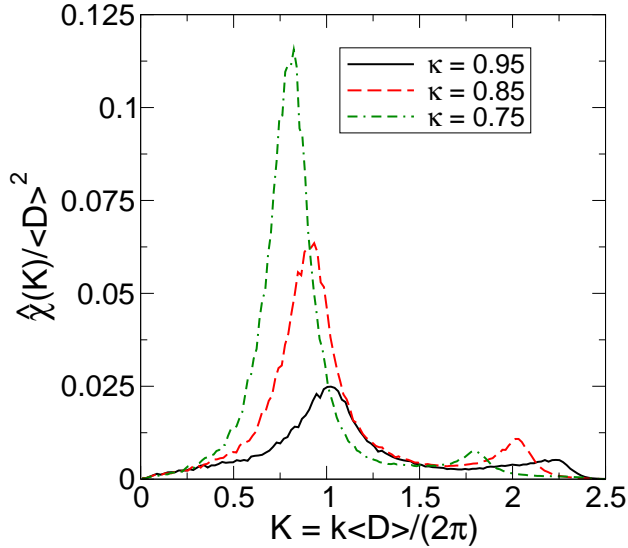}
\caption{(Color online)
Spectral densities upon shrinking the particles in the binary 
MRJ packings at fixed size ratio $\beta = R_{\text{large}}/R_{\text{small}}$.  
The parameter $\kappa = R_{\text{shrink}}/R_{\text{MRJ}}$
is the scaling factor.}\label{fignineA}
\end{figure}

To motivate our discussion of the void space and its fundamental importance to polydisperse MRJ packings, we mention a peculiar
property of hyperuniformity in these packings.
Namely, one can maintain hyperuniformity in polydisperse MRJ hard-particle systems 
even upon shrinking the particles at a fixed size ratio so long as one does not affect the underlying 
statistical distribution of the point process; this effect is illustrated in Figure \ref{fignineA}.  
To understand this behavior, we note that the leading-order term governing the expansion 
of the Fourier transform of the particle indicator function $\hat{m}(k; R)$ is the volume of the particle $\pi R^2$, 
meaning that for small wavenumbers we have (in the case of the binary packings)
\begin{equation}\label{fiftysevenA}
\hat{\chi}(\mathbf{k}) \sim \frac{1}{V}\left\lvert \pi R_{\text{small}}^2 \sum_{j=1}^{N_{\text{small}}} \exp(i \mathbf{k}\cdot \mathbf{r}_j) 
+ \pi R_{\text{large}}^2 \sum_{\ell=1}^{N_{\text{large}}} \exp(i \mathbf{k}\cdot \mathbf{r}_\ell)\right\rvert^2,
\end{equation}
which, upon rescaling $R_i \rightarrow \kappa R_i$ for $\kappa <1$, suggests
\begin{equation}
\hat{\chi}_{\text{shrink}}(\mathbf{k}) \sim \kappa^4 \hat{\chi}_{\text{MRJ}}(\mathbf{k}) \qquad (k \rightarrow 0).
\end{equation}
Therefore, hyperuniformity of the MRJ binary packing is not lost when performing this scaling operation.  
Physically, the small-wavenumber region of the spectral density effectively homogenizes 
the medium due to the coupling between the wavenumber and the particle radius in the indicator function $\hat{m}(k; R)$, 
meaning that this region is not affected by changes at the boundaries of the particles so long
as the underlying statistics of the point pattern generating the medium remain constant.  
This observation suggests that a saturated and strictly jammed sphere packing extends naturally to an uncountably infinite family 
of hyperuniform heterogeneous media related by an appropriate scaling parameter.  
One can map this behavior directly using the result \eqref{qform}, which indicates that the 
spectral density at small wavenumbers defines a quadratic form in the particle radii when the 
underlying point pattern of the disk centers is held fixed; the corresponding curve is plotted 
in Figure \ref{qformfig}.
Furthermore, this scaling has the effect of deforming the void space surrounding the particles in a \emph{uniform} manner 
that preserves the regularity of the pore distribution even upon breaking both the jamming and saturation constraints.  
The medium therefore is able to retain information from the strictly jammed configuration 
even upon relaxing the sizes of the inclusions.  
Additionally, it follows from the quadratic form \eqref{qform} that \emph{non-uniform} changes
in the particle radii inherently break the hyperuniformity of the packing.

This behavior has important implications for MRJ packings.  First, this effect is geometric in origin and depends on
the explicit inclusion of particle shape information to appropriately balance the partial structure factors of the 
packing.  This observation immediately implies that ``point'' information contained in the particle centers
is not sufficient to describe the system.  Second, the presence of interparticle contacts is not essential for
the onset of hyperuniformity and QLR correlations.  This subtle point, though easy to appreciate mathematically,
is highly nontrivial from a physical perspective.  It implies that on the \emph{global} scale of the microstructure, the 
space is regularized in such a way that hyperuniformity is preserved upon making uniform scaling deformations of the 
particle phase.  Importantly, since the spectral density is a descriptor of both the particle and the void phases, 
this homogenization is invariant to the choice of reference phase.  Note in particular that upon breaking the contact 
network of the MRJ packing, the void space can become \emph{connected} throughout the space, implying that
the spatial statistics of the microstructure must account for correlations over large length scales.  It is for these reasons
that we turn our attention to the void-space distribution in the following sections.     

Note that our scaling analysis does not apply if the particle radii are 
uniformly increased since breaking the impenetrability constraint implies that higher-order microstructural 
information is necessary to characterize the medium, and \eqref{fiftysevenA} is no longer valid.  Additionally, 
if the particles are shrunk to the point that the small particle radius vanishes, then the packing will no longer
be hyperuniform 
since important shape information about the system has been lost.  In terms of the void space, the distribution of 
pore sizes will be skewed toward higher values due to the sudden appearance of ``holes'' in the microstructure, 
corresponding to the lost small particles, thereby de-regularizing the microstructure and breaking hyperuniformity.

\section{Characterization of the pore-size distribution}

\subsection{Numerical evaluation of pore-size statistics}

It is clear from the discussion above that the void phase surrounding 
the disk inclusions of the binary MRJ packings plays 
a significant role in characterizing local fluctuations in the medium.  
Indeed, we argue that the conditions of strict jamming and saturation place strong
constraints on the distribution of the pore sizes, effectively regularizing 
the local structure around each disk such that the system is 
hyperuniform with respect to local-volume-fraction fluctuations even though 
the disk centers do not consitute a hyperuniform point pattern.  
To support our arguments concerning the void space, we 
have quantified the size of the available void space using the 
so-called \emph{complementary pore-size cumulative distribution function} 
$F(\delta)$ \cite{torquato2002rhm}, which represents the 
fraction of the void space external to the inclusion phase with a pore radius 
larger than $\delta$.  Equivalently, if we define $P(\delta)$ 
as the probability density that a randomly chosen point in the void space lies 
within $\delta$ and $\delta + d\delta$ from the \emph{nearest}
point on the void-inclusion interface, then:
\begin{equation}
F(\delta) = \int_{\delta}^{+\infty} P(\Delta) d\Delta.
\end{equation}

\begin{figure}[!tp]
\centering
\includegraphics[width=0.50\textwidth]{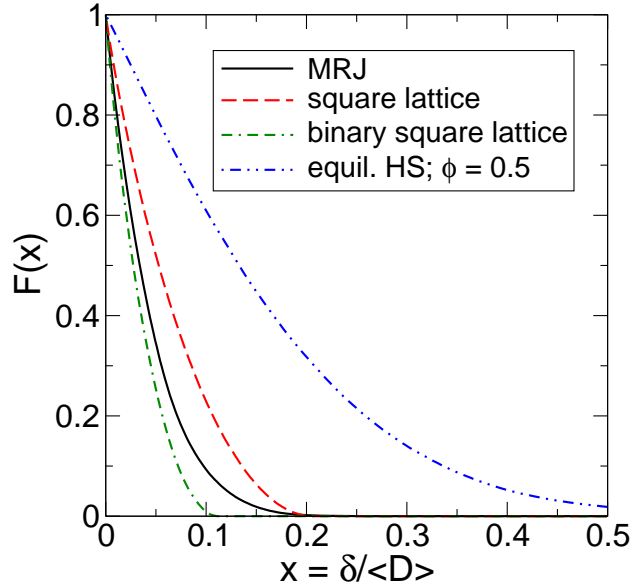}
\caption{(Color online)  Cumulative pore-size distributions $F$ for the binary 
MRJ packing, the square lattice $\mathbb{Z}^2$, 
a saturated binary variant of $\mathbb{Z}^2$, and 
a system of equilibrium hard spheres with volume fraction $\phi = 0.5$.}\label{porefig}
\end{figure}

Figure \ref{porefig} shows the cumulative pore-size distribution function for the 
binary MRJ packings along with the corresponding results
for the integer and binary integer packings and a system of equilibrium 
monodisperse hard disks at volume fraction $\phi = 0.5$.  The equilibrium hard disk
packing is neither saturated nor jammed; it is also known that this system is not 
hyperuniform \cite{ToSt03}.  As a result, it possesses 
a broad distribution of pore sizes, and the probability of finding large pores 
(i.e., on the order of a particle diameter) is 
nonvanishing.  In contrast, both the integer packing and its binary variant possess 
narrow pore-size distributions with compact supports, which result from 
the close-packed nature of the packings.  Interestingly, Figure \ref{porefig} indicates 
that the binary MRJ packing also possesses a narrow pore-size distribution 
that essentially falls between the square and binary square packings.    
This observation suggests that the void space
is highly constrained by the condition of strict jamming and almost regular in its distribution.

One can directly compute the moments $\langle \delta^n \rangle$ of the pore size density from the cumulative distribution $F(\delta)$ using the relation \cite{torquato2002rhm}:
\begin{equation}
\langle \delta^n\rangle = n\int_{0}^{+\infty} \delta^{n-1} F(\delta) d\delta.
\end{equation}
Lower-order moments of the pore-size density arise in bounds on the mean survival times and principal relaxation times of heterogeneous materials and are thus 
important descriptors of microstructures \cite{Pr63, ToAv91} (see Section V.B below).  Table \ref{tableone} provides the first moments and standard deviations of 
the pore size for the
systems mentioned in Figure \ref{porefig}.  The average pore sizes for the hyperuniform systems are significantly smaller than for the equilibrium hard disk packing. 
In fact, the pore size distribution of the binary MRJ packing appears to be more localized about its mean than even the integer packing.  We conclude from this 
information that the available void space in the binary MRJ disk packing is sufficiently restricted such that local volume fraction fluctuations 
decay faster than the volume of an observation window, thereby reflecting the underlying regularity of the pore space.  
\begin{table}[!tp]
\caption{First moments and standard deviations of the pore size for distributions of hard disks.}\label{tableone}
\begin{ruledtabular}
\begin{tabular}{c c c}
 & $\langle\delta\rangle/\langle D\rangle$ & $\sigma_{\delta}/\langle D\rangle$\\
\hline
binary MRJ & 0.0438 & 0.0380\\
$\mathbb{Z}^2$ & 0.0626 & 0.0480\\
binary $\mathbb{Z}^2$ & 0.0332 & 0.0258\\
equil. HS ($\phi = 0.5$) & 0.1612 & 0.1267
\end{tabular}
\end{ruledtabular}
\end{table}

\subsection{Bounds on the pore size distribution}

It is important to note that the pore-size distribution for a
hyperuniform hard-particle packing with volume fraction $\phi$ must asymptotically 
decay faster than the corresponding distribution for a non-hyperuniform medium with the same 
one-point statistics; 
this claim is a generalization of an argument by Gabrielli and Torquato \cite{GaTo04}.  
Since non-hyperuniform media are by definition ``Poisson-like,'' it suffices then to consider the 
pore-size distribution for a polydisperse system of fully penetrable disks.  
The cumulative pore-size distribution $F_P(\delta)$ for such a system is known analytically 
\cite{St95, torquato2002rhm} and is given by
\begin{equation}
F_P(\delta) = (1/\phi_1)\exp\left[-\rho \langle v(\delta + R)\rangle_R\right],
\end{equation}
where the angular brackets denote an average over the disk radii and $\phi_1$ is the volume 
fraction of the matrix phase.  
We emphasize in these calculations that we are considering the pore-size distribution for a 2D
\emph{Poisson-distributed} 
heterogeneous medium with volume fraction $\phi$ that is the \emph{same as a reference 
hyperuniform system}.  
However, equivalence of the one-point statistics does not imply that the number densities 
are the same between the systems.  
To account for this discrepancy, we write $\rho = \eta/(\pi \langle R^2\rangle_R) 
= \ln(1/\phi_1)/(\pi\langle R^2\rangle_R)$, which implies
\begin{align}
F_P(\delta) = \exp\left[-(s\langle D\rangle_R/\phi_1)\tilde{\delta}^2 
- (s\langle D\rangle_R/\phi_1) \tilde{\delta}\right]\label{twentysixa},
\end{align}
where $s = \rho \phi_1 \langle s(R)\rangle_R$ is the 
specific surface of a Poisson-distributed medium and $s(R)$ is the surface area of a
$d$-dimensional sphere of radius $R$. 
Note that we have introduced the length scale $\langle D\rangle_R$ 
and reduced variable $\tilde{\delta} = \delta/\langle D\rangle_R$ in \eqref{twentysixa}.  
Letting $b = s\langle D\rangle_R/\phi_1$, we find the following upper 
bounds for the first and second moments of the pore size density:
\begin{align}
\langle\tilde{\delta}\rangle_{\text{UL}} &= \frac{1}{2}\sqrt{\frac{\pi}{b}} 
\exp\left[b/4\right] \text{erfc}\left(\sqrt{b}/2\right)\label{twentysevena}\\
\langle\tilde{\delta}^2\rangle_{\text{UL}} &= 
\frac{1}{b}\left(1-b\langle\tilde{\delta}\rangle_{\text{UL}}\right)\label{twentyeighta}.
\end{align}

The above Poisson bound will be rigorously true for the full pore-size distribution 
of our polydisperse MRJ packings based solely on the presence
of hyperuniformity.  However, we can significantly tighten the bounds of the moments of the pore-size distribution 
for binary MRJ packings based on our analysis below of the local
voids.  Specifically, since the distribution of voids within the MRJ packings is dominated 
by three- and four-particle loops, the average pore size 
must be less than the corresponding average pore size of a saturated integer lattice.  
This effect is explicitly shown in Figure \ref{porefig}, where it is clear
that the integer lattice provides a good approximate upper bound for the full pore-size 
distribution with the exception of the tail, due to the presence 
of higher-order loops in the MRJ packing.  Since this tail must decrease faster than 
exponentially with a cut-off at the small-particle radius (because of saturation),
we therefore have the following tighter upper bounds on the first and second moments 
of the pore size distribution:
\begin{align}
\langle\tilde{\delta}\rangle &\leq \langle\tilde{\delta}\rangle_{\mathbb{Z}^2}\\
\langle\tilde{\delta}^2\rangle &\leq \langle\tilde{\delta}^2\rangle_{\mathbb{Z}^2}.
\end{align}

\begin{figure}[!t]
\centering
\includegraphics[width=0.5\textwidth]{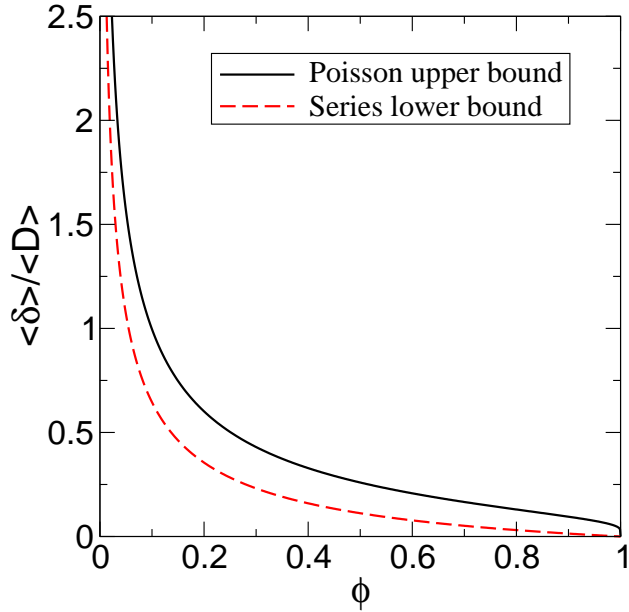}
\caption{(Color online)  Upper and lower bounds on the mean pore size applicable for binary hyperuniform 
heterogeneous media at volume fraction $\phi$ with the same composition as the MRJ binary disk 
packings studied here.}\label{porebnds}
\end{figure}

It is also possible to find a simple lower bound on the pore-size distribution that is 
applicable for arbitrary heterogeneous media (hyperuniform or not).  
Specifically, we utilize the following series representation for the cumulative pore-size 
distribution \cite{torquato2002rhm}:
\begin{equation}\label{thirtya}
F(\delta) = \frac{1}{\phi_1}\left[1+\sum_{k=1}^{+\infty} 
(-\rho)^k \left(\frac{1}{\Gamma[k+1]}\right)\int \left\langle g_k(\mathbf{r}^k; R^k) 
\prod_{j=1}^k m(\lVert \mathbf{x}-\mathbf{r}_j\rVert; \delta+R_j)\right\rangle_{R^k} d\mathbf{r}_j\right].
\end{equation}
It is well-known that keeping terms up to order $\rho^k$ in \eqref{thirtya} places 
an upper bound on $F(\delta)$ for $k$ even 
and enforces a lower bound for $k$ odd \cite{torquato2002rhm}.  Therefore, by 
expanding the series to order $\rho$ 
we obtain the following lower bound on $F(\delta)$:
\begin{equation}
F(\delta) \geq \text{max}\left\{\frac{1}{\phi_1}\left[1-\rho \langle v(\delta + R)\rangle_R\right], 
0\right\}\label{thirtyonea};
\end{equation}
the max operation in \eqref{thirtyonea} enforces the trivial lower bound $F(\delta) \geq 0$.  
Note that \eqref{thirtyonea}
 is equivalent to the $\mathcal{O}(\rho)$ expansion of the pore-size distribution for a 
 Poisson-distributed medium; 
 however, the number density (and therefore the specific surface) of the system is not 
 necessarily the same as the Poisson medium.  
 We will now assume that our hyperuniform reference medium consists of impenetrable 
 spheres with volume fraction $\phi = \rho \langle v(R)\rangle_R$; 
 we also remark that the definition of the specific surface for impenetrable spheres 
 $s = \rho\langle s(R)\rangle_R$ is 
 different than for a fully-penetrable system.  It now follows that
\begin{align}
F(\delta) &\geq \text{max}\left\{1-b \tilde{\delta}^2 - b\tilde{\delta}, 0\right\}\label{thirtyfivea},
\end{align}
where $b = s\langle D\rangle_R/\phi_1$ as in the Poisson case. 
The lower bound \eqref{thirtyfivea} first reaches zero at
\begin{equation}
\tilde{\delta^*} = \frac{-1 + \sqrt{1 + 4/b}}{2}.
\end{equation}
 We thus obtain the following lower bounds on the first and second moments of the pore size density
\begin{align}
\langle \tilde{\delta}\rangle_{\text{LL}} &= \tilde{\delta^*}- \frac{b \tilde{\delta^*}^2}{2} - \frac{b 
\tilde{\delta^*}^3}{3}\label{thirtysixa}\\
\langle \tilde{\delta}^2\rangle_{\text{LL}} &= \tilde{\delta^*}^2 - \frac{2 b \tilde{\delta^*}^3}{3} 
- \frac{b \tilde{\delta^*}^4}{2}\label{thirtysevena}.
\end{align}
Figure \ref{porebnds} plots the upper and lower bounds \eqref{twentysevena} and \eqref{thirtysixa} on 
the mean pore size for hyperuniform binary heterogeneous media at volume fraction $\phi$.  
We emphasize that these bounds account 
only for the hyperuniformity of the packings and thus place constraints on the pore-size distributions 
of \emph{any} binary hyperuniform
hard-particle packing.  

For our binary MRJ packings, the specific surface is
\begin{align}
s_{\text{MRJ}}\langle D\rangle_R/\phi_1 &= (\phi/\phi_1) \langle D\rangle_R^2/\langle R^2\rangle_R = 
(4\phi/\phi_1)\left[\frac{(\gamma_{\text{small}} + \gamma_{\text{large}} \beta)^2}{\gamma_{\text{small}} 
+ \gamma_{\text{large}} \beta^2}\right] \approx 21.6917.
\end{align}
The specific surface for the equivalent Poisson-distributed medium is
\begin{equation}
s_P \langle D\rangle_R/\phi_1 = -\ln(\phi_1) \langle D\rangle_R^2/\langle R^2\rangle_R \approx 7.34037;
\end{equation}
as expected, this value is less than the corresponding result for the impenetrable case.   
Using these parameters, we collect in Table \ref{tableoneA} the bounds on the pore size of our binary 
MRJ disk packings.
We immediately notice that the numerical values are well within the bounds above; 
it follows that the one-point statistics of the medium do not place strong constraints on the distribution of 
the void space.  
Furthermore, this evidence suggests hyperuniformity strongly regularizes the pore space compared to 
the Poisson-distributed system.
\begin{table}
\caption{Bounds on the moments of the pore-size distribution for binary MRJ packings 
of hard disks.}\label{tableoneA}
\begin{tabular}{c c c c c}
\hline\hline
~ & ~~~& $\langle\tilde{\delta}\rangle$ &~~~ & $\langle\tilde{\delta}^2\rangle$\\
\hline
Poisson upper bound  & & 0.1135 & & 0.0227\\ 
$\mathbb{Z}^2$ upper bound & & 0.0626 & & 6.219$\times 10^{-3}$\\
Series lower bound & & 0.0224 & & 6.635$\times 10^{-4}$\\
MRJ binary packing & & 0.0438 & & 3.362$\times 10^{-3}$\\
\hline\hline
\end{tabular}
\end{table}

It is interesting to note that the series lower bound provides a better estimate 
of the mean pore size than the Poisson upper bound.  
This observation is reasonable since the compact support of the lower bound 
more closely matches the behavior of the actual pore-size distribution, 
which also has compact support due to the saturation of the packing.  However, 
it is unclear if the constraint of saturation can be relaxed 
while still maintaining hyperuniformity in the medium. One can conjecture that 
so long as the probability of finding a large pore 
decays faster than the corresponding behavior for a Poisson pattern, then hyperuniformity holds.   

Although our 
results provide a quantitative basis for understanding the appearance of hyperuniformity in 
polydisperse MRJ packings,
a complete explanation for the linear scaling of the small-wavenumber region of the spectral 
density is still not apparent.  
Again, we emphasize that this problem is inherently non-local, and its solution must 
account for the presence of QLR correlations
within the packing.  Such correlations are difficult to discern with the distribution 
$F(\delta)$ that we have presented here, which 
is essentially a one-point descriptor of heterogeneous media \cite{torquato2002rhm}.  

In the following section, we characterize the allowable $n$-particle loops 
within a binary MRJ packing.
This analysis suggests 
that the variance in pore sizes is bounded by the strict jamming
of the packings, supporting our argument these constraints are sufficient to 
induce hyperuniformity.  

\section{Characterization of $n$-particle loops in MRJ binary disk packings}

The preceding discussion shows 
how the void space is a \emph{fundamental} descriptor of a two-phase random heterogeneous medium.
As we have seen, fluctuations in the local volume fraction contain 
information about the void space and therefore provide a more complete picture of hyperuniformity 
in heterogeneous media.  
Here we quantitatively characterize the voids in
MRJ binary disk packings and argue that strict jamming restricts the variance 
of pore sizes. 

An \emph{$n$-particle void} 
is associated with a loop of $n$ contacting disks (an
\emph{$n$-particle loop}) in which each disk only contacts two neighbors.
These loops are defined topologically, meaning that their identification is invariant to 
local shears of the particle contacts.
The smallest loops contain three mutually contacting particles.
The constraint of saturation places an upper bound on the number of 
particles in a loop, and the largest
loop we find in the MRJ binary disk packings contains 6 particles. 
The number of $n$-particle
loops in a packing decreases rapidly as $n$ increases.

The area of the void associated with a loop can be rigorously
computed by substracting the area of particles falling into the
polygon constructed by connecting the centers of the disks in the
loop. We note that, except for the 3-particle loop, the
polygons associated with other $n$-loops ($n \ge 4$) can be continuously deformed 
while maintaining the contacts between the particles. Such deformations
change the area of the voids associated with the loop. 
The rigidity of MRJ packings also strongly constrains the number of
particles in a loop. In the MRJ packings of binary disks, the
majority of the voids involve 3-particle loops, which 
form the ``backbone'' of the network. Certain 4-particle loops can
be observed at the effective grain boundaries between disks with different
sizes. A large portion of these 4-particle loops are very
distorted with a void area almost equal to that
associated with two 3-particle loops. Loops with more particles
are rare in the packings, and the observed ones are all
strongly distorted with void areas almost equal to those
associated with 3- or 4-particle loops. Therefore, we only
discuss the voids associated with 3-
and 4-particle loops here, focusing on the maximum possible void areas associated with these
loops.

\begin{figure}
\centering
\includegraphics[width=0.50\textwidth]{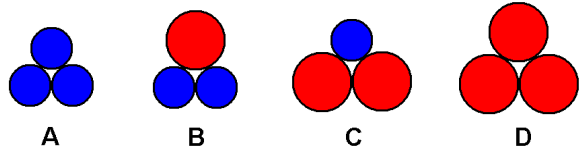}
\caption{(Color online)  The four distinct 3-particle loops in a binary circular
disk packing.} \label{fig3loop}
\end{figure}
There are four distinct
3-particle loops in the binary disk packings as shown in Figure \ref{fig3loop}. Defining the radii
of small and large disks be $R_1$ and $R_2$, respectively, the
$n$-particle
void areas $\lambda_{n\alpha}$, normalized by the volume of a small particle, are
$\lambda_{3A} \approx 0.0513$, $\lambda_{3B} \approx
0.0631$, $\lambda_{3C} \approx 0.0790$ and $\lambda_{3D} \approx
0.1060$ for a size ratio $\beta = R_2/R_1 = 1.4$.  
It can be seen clearly that the void
areas constitute a small fraction of the small particle area. 
Since these 3-particle voids dominate the packing, it is reasonable to 
conclude that the regularity of these local clusters enforces 
hyperuniformity on the medium despite the nonuniform 
distribution of the sphere centers.  

\begin{figure}
\centering
\includegraphics[width=0.50\textwidth]{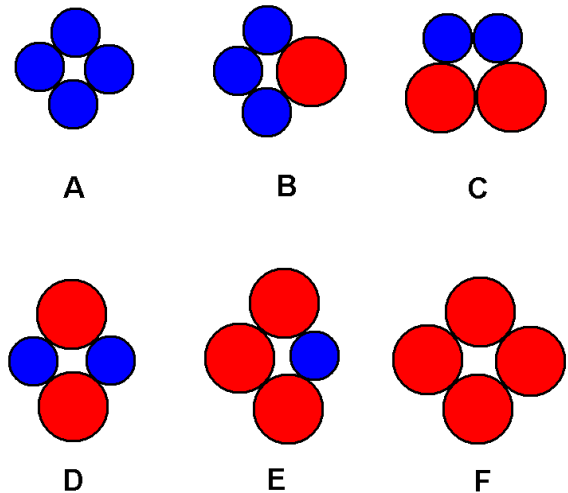}
\caption{(Color online)  The six distinct families of 4-particle loops in a binary
circular disk packing.} \label{fig4loop}
\end{figure}
There are total of six distinct types 
of 4-particle
loops (see Figure \ref{fig4loop}) in the binary disk packings, 
each associated with a void that
can possess a spectrum of shapes and sizes by distorting the
quadrangle formed by the centers of the disks. Although most of
the void areas associated with these highly distorted 4-particle loops 
are almost equal to the area associated with two
3-particle loops, there are still relatively large voids. 
The maximum normalized void areas are given by
$\lambda_{4A} \approx 0.2732$,
$\lambda_{4B} \approx 0.3208$, $\lambda_{4C} \approx 0.3934$,
$\lambda_{4D} \approx 0.3782$, $\lambda_{4E} \approx 0.4498$ and
$\lambda_{4F} \approx 0.5355$. 
These values are relatively large but more tightly distributed around the 
mean when
compared to the normalized areas for 3-particle voids. As
a result, they can be considered as an estimate of the
upper bound on the degree of local inhomogeneity that can be
consistent with hyperuniformity of the packings. 


\section{Concluding remarks}

We have provided a detailed study of local-volume-fraction fluctuations in MRJ packings of polydisperse hard disks and have shown that 
these systems are hyperuniform with quasi-long-range correlations.  Our results strongly suggest that these QLR correlations are a signature of 
MRJ hard-particle packings, in contrast to previous misconceptions in the literature \cite{XuCh10}.  Although it is true that the structure factors
for these systems do not vanish at small wavenumbers except in the monodisperse limit, we have shown that the more appropriate 
structural descriptor of MRJ packings is the spectral density, which accounts appropriately for the shape information of the particles.  Our
work therefore generalizes the Torquato-Stillinger conjecture for hyperuniform point patterns by suggesting that all saturated, strictly-jammed 
sphere packings are hyperuniform with respect to local-volume-fraction fluctuations.  
Furthermore, MRJ sphere packings are expected to exhibit quasi-long-range pair correlations
scaling as $r^{-(d+1)}$ in $d$ Euclidean dimensions.
Importantly, this generalization contains previously-published 
results for monodisperse MRJ packings \cite{DoStTo05} 
as the special case where the particle shape information can be rigorously separated from the
``point'' information of the particle centers.  

Based on the observations that local-volume-fraction fluctuations are invariant to 
the choice of reference phase and that one can maintain hyperuniformity
in MRJ polydisperse packings under a uniform scaling deformation of the particles, we have 
argued that the onset of hyperuniformity and quasi-long-range
correlations results from the homogenization of the void space external to the particles.  The conditions of saturation and strict jamming limit the 
sizes and shapes of the local voids, which are completely determined by the contact network of particles.  
Using a local analysis of the voids and rigorous 
bounds on the pore-size distribution, we have argued that the void space is highly constrained by strict jamming, 
thereby suppressing infinite-wavelength 
local-volume-fraction fluctuations.  Furthermore, we suggest that the presence of quasi-long-range correlations reflects the inherent structural 
correlations from the contact network between void shapes.  Specifically, saturation and strict jamming of the packings compete with the 
maximal randomness of the particle distributions to drive the small-wavenumber region of the spectral density to its smallest integer value.  

Although our work has addressed important problems related to the structural properties of MRJ hard-particle packings, a number of unanswered 
questions remain.  First, a rigorous foundation for the observed linear scaling in the small-wavenumber region of the spectral density is still missing.  
This problem is immensely difficult to handle theoretically because the problem is inherently long-ranged, and any local analysis of the 
MRJ structure will therefore be unable to account correctly for this behavior \cite{ToSt10}.  Second, although we have considered only polydisperse
MRJ packings of $d$-dimensional spheres, our results are expected to hold more generally for strictly jammed packings of hard particles of 
arbitrary geometry.  Indeed, our arguments concerning the void space of an MRJ packing are easily extended to include these more general cases, and 
in a companion paper we will provide direct evidence that MRJ packings of hard ellipses and superdisks \cite{ZaJiTo10} are also hyperuniform with signature
quasi-long-range correlations indicated by linear scaling of the small-wavenumber region of the spectral density.  These results will suggest a remarkably strong extension of the Torquato-Stillinger conjecture,
namely that all maximally random strictly jammed saturated packings of hard particles, including 
those with size- and shape-distributions, are hyperuniform with universal quasi-long-range 
correlations [via the two-point probability function $S_2(r)$] scaling asymptotically as $r^{-(d+1)}$.  

We mention that recent, independent work, appearing
in preprint form shortly after our own manuscript was posted, 
\cite{BeChCoDa11} has shown that hyperuniformity
and vanishing infinite-wavelength local-volume-fraction fluctuations in saturated MRJ 
packings of polydisperse disks are consistent with certain fluctuation-response relations
involving a generalized ``compressibility.''  However, this work does not address 
the appearance of quasi-long-range pair correlations in MRJ hard-particle packings and 
does not consider MRJ packings of nonspherical particles as in our companion paper \cite{ZaJiTo10}.  

Our arguments suggest in particular that certain quantum 
many-body systems and cosmological structures with the same linear small-wavenumber scaling 
in the structure factor \cite{ToScZa08, ReCh67, Pe93} 
are statistically ``rigid'' in the sense that their microstructures 
are effectively homogeneous over large length scales with vanishing infinite-wavelength 
number density fluctuations.  These unique features are inherently linked to the \emph{structural}
properties of the system, independent of the physical model itself.   

Finally, we note that the systems examined in this paper are constrained to be both saturated and strictly jammed.  Saturation of the packings is responsible for 
enforcing compact support in the pore-size distribution function and therefore plays an important role in regularizing the void space surrounding the jammed disks.  
However, it is not clear if saturation is a necessary condition to ensure hyperuniformity in strictly jammed heterogeneous media; this situation corresponds to 
a weakening of the Torquato-Stillinger conjecture.  Namely, what conditions must a strictly jammed but unsaturated packing of hard spheres meet in order to be hyperuniform?    
We note that the event-driven molecular dynamics algorithm used here to generated the binary MRJ disk packings inherently precludes the presence of 
arbitrarily large ``holes'' within the packing.  However, it does produce a small concentration ($\sim$2.5\%) of ``rattler'' particles, which are particles that are free to move 
within some small caged region of the packing.  Removal of such particles is known to break hyperuniformity of the medium even though the strict
 jamming of the surrounding structure
remains \cite{DoStTo05}; therefore, strict jamming alone is not sufficient to induce hyperuniformity if large holes are ``common enough'' in the statistics of the microstructure.  
This scenario corresponds to skewing the pore-size distribution and thereby de-regularizing the void space.  

\begin{figure}[!t]
\centering
\includegraphics[width=0.50\textwidth]{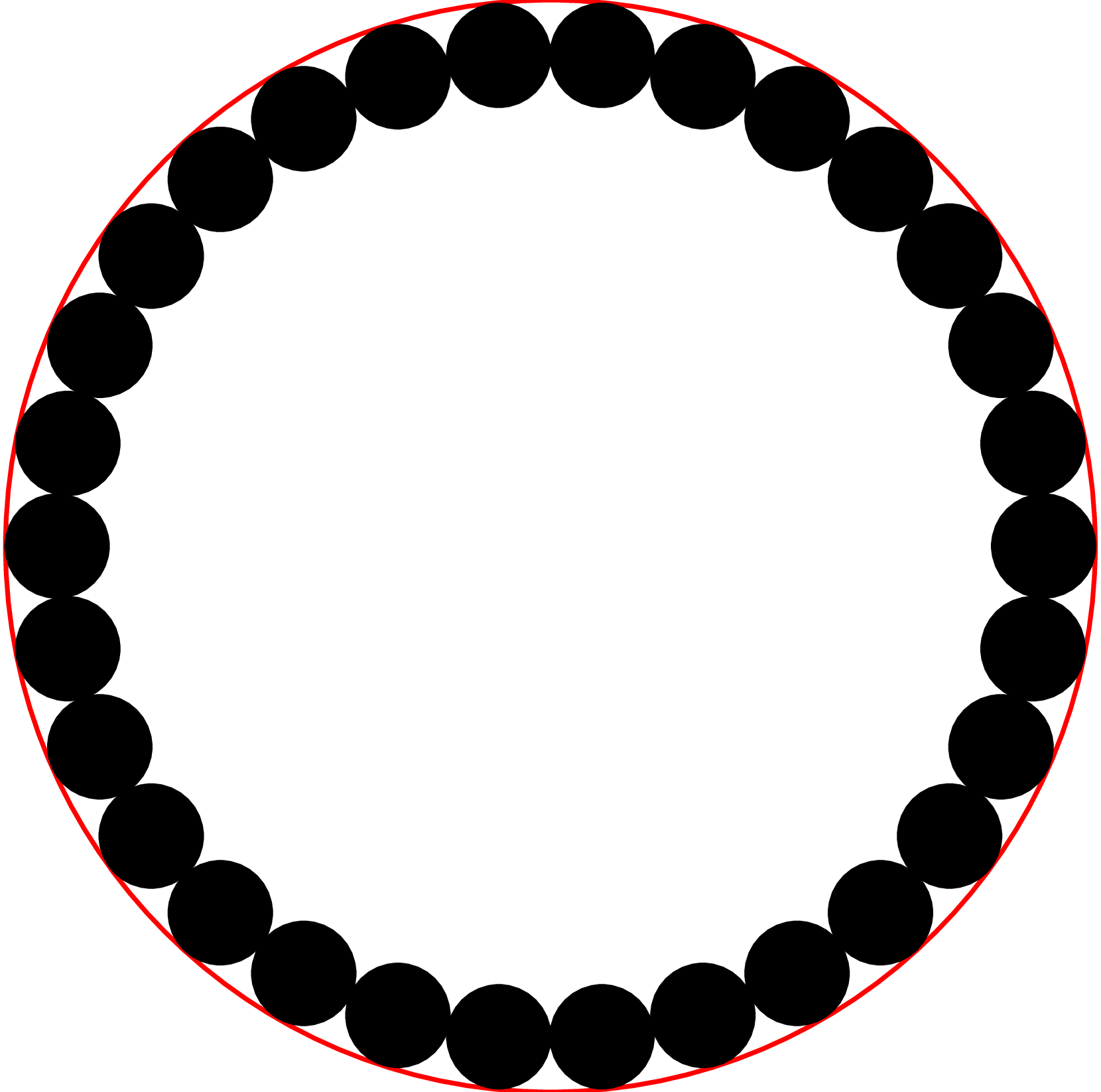}
\caption{(Color online)  A jammed configuration of particles (up to
rotations) surrounding a circular void.  By replacing the wall with a 
close-packed collection of particles, one can construct 
strictly jammed packings in the plane with pores of arbitrarily large size.}\label{hole}
\end{figure}

It is indeed possible to construct strictly
jammed packings of two-dimensional
disks with a hole of arbitrarily large size by starting with a ring of particles encompassing a large pore; this ring can be jammed by surrounding it with a close-packed 
collection of particles that approximates an impenetrable ``wall'' (see Figure \ref{hole}). 
One is then free to construct any strictly jammed system of particles outside the hole \cite{FN8}.  
If such holes are sufficiently rare, then it is possible that the system may still be hyperuniform since the pores do not become very large on average.

\begin{acknowledgments}
This work was supported by the National Science Foundation under Grants DMS-0804431 and DMR-0820341.
\end{acknowledgments}


\begin{thebibliography}{48}
\bibitem{FN1}
  A packing in which no space is available to insert an additional particle without violating
  the impenetrability conditions is called saturated.
\bibitem{FN2}
  The converse statement is not true, meaning that hyperuniformity does not imply
  strict jamming.  The square lattice in two dimensions is one example of 
  a hyperuniform packing that is not strictly jammed.
\bibitem{FN4}
  The spectral density possesses the properties of a probability density 
  when normalized by $\chi(\mathbf{0}) = \phi(1-\phi)$; the normalized 
  function $\tilde{g}(\mathbf{k}) = \hat{\chi}(\mathbf{k})/\chi(\mathbf{0})$
  is also known as the spectral density.
\bibitem{FN5}
Unless otherwise noted, we will use subscripts on the angular
brackets to distinguish such radial averages from ensemble averages.
\bibitem{FN6}
Although we only consider packings in 2D for this study, our results are
expected to apply in higher dimensions as well.  We have already seen that
our methodology contains known results for 3D monodisperse MRJ
sphere packings as a special case, and the arguments we make 
in this section concerning the constrained void space extend directly
to higher-dimensional packings.
\end{thebibliography}

\end{document}